\documentclass[twocolumn]{aastex631}

\usepackage{subfig}
\usepackage{upgreek}
\usepackage{CJK}
\usepackage{lineno}

\shorttitle{\textit{JWST}-SUSPENSE: Elemental abundances}
\shortauthors{Aliza G. Beverage}

\begin{document}
\title{Carbon and Iron Deficiencies in Quiescent Galaxies at $z=1-3$ from \textit{JWST}-SUSPENSE:\\
Implications for the Formation Histories of Massive Galaxies}

\correspondingauthor{Aliza Beverage}
\email{abeverage@berkeley.edu}

\author[0000-0002-9861-4515]{Aliza G. Beverage}
\affiliation{Department of Astronomy, University of California, Berkeley, CA 94720, USA}

\author[0000-0001-7540-1544]{Martje Slob}
\affiliation{Leiden Observatory, Leiden University, P.O. Box 9513, 2300 RA Leiden, The Netherlands}

\author[0000-0002-7613-9872]{Mariska Kriek}
\affiliation{Leiden Observatory, Leiden University, P.O. Box 9513, 2300 RA Leiden, The Netherlands}

\author[0000-0002-1590-8551]{Charlie Conroy}
\affiliation{Center for Astrophysics \textbar\ Harvard \& Smithsonian, Cambridge, MA, 02138, USA}

\author[0000-0001-6813-875X]{Guillermo Barro}
\affiliation{Department of Physics, University of the Pacific, Stockton, CA 90340 USA}

\author[0000-0001-5063-8254]{Rachel Bezanson}
\affiliation{Department of Physics and Astronomy and PITT PACC, University of Pittsburgh, Pittsburgh, PA 15260, USA}

\author[0000-0003-2680-005X]{Gabriel Brammer}
\affiliation{Cosmic Dawn Center (DAWN), Denmark}
\affiliation{Niels Bohr Institute, University of Copenhagen, Jagtvej 128, DK2200 Copenhagen N, Denmark}

\author[0000-0003-2251-9164]{Chloe M. Cheng}
\affiliation{Leiden Observatory, Leiden University, P.O. Box 9513, 2300 RA Leiden, The Netherlands}

\author[0000-0002-2380-9801]{Anna de Graaff}
\affiliation{Max-Planck-Institut f\"ur Astronomie, K\"onigstuhl 17, D-69117, Heidelberg, Germany}

\author[0000-0003-4264-3381]{Natascha M. F\"orster Schreiber}
\affiliation{Max-Planck-Institut f\"ur extraterrestrische Physik, Giessenbachstrasse 1, D-85748 Garching, Germany}

\author[0000-0002-8871-3026]{Marijn Franx}
\affiliation{Leiden Observatory, Leiden University, P.O. Box 9513, 2300 RA Leiden, The Netherlands}

\author[0000-0002-5337-5856]{Brian Lorenz}
\affiliation{Department of Astronomy, University of California, Berkeley, CA 94720, USA}

\author[0000-0001-5175-939X]{Pavel E. Mancera Pi\~na}
\affiliation{Leiden Observatory, Leiden University, P.O. Box 9513, 2300 RA Leiden, The Netherlands}

\author[0000-0001-9002-3502]{Danilo Marchesini}
\affiliation{Department of Physics \& Astronomy, Tufts University, MA 02155, USA}

\author[0000-0002-9330-9108]{Adam Muzzin}
\affiliation{Department of Physics and Astronomy, York University, 4700 Keele Street, Toronto, Ontario, ON MJ3 1P3, Canada}

\author[0000-0001-7769-8660]{Andrew B. Newman}
\affiliation{Observatories of the Carnegie Institution for Science, 813 Santa Barbara St., Pasadena, CA 91101, USA}

\author[0000-0002-0108-4176]{Sedona H. Price}
\affiliation{Department of Physics and Astronomy and PITT PACC, University of Pittsburgh, Pittsburgh, PA 15260, USA}

\author[0000-0003-3509-4855]{Alice E. Shapley}
\affiliation{Department of Physics \& Astronomy, University of California, Los Angeles, CA 90095, USA}

\author[0000-0001-7768-5309]{Mauro Stefanon}
\affiliation{Departament d'Astronomia i Astrof\`isica, Universitat de Val\`encia, C. Dr. Moliner 50, E-46100 Burjassot, Val\`encia,  Spain}
\affiliation{Unidad Asociada CSIC "Grupo de Astrof\'isica Extragal\'actica y Cosmolog\'ia" (Instituto de F\'isica de Cantabria - Universitat de Val\`encia)}

\author[0000-0002-1714-1905]{Katherine A. Suess}
\altaffiliation{NHFP Hubble Fellow}
\affiliation{Kavli Institute for Particle Astrophysics and Cosmology and Department of Physics, Stanford University, Stanford, CA 94305, USA}

\author[0000-0002-8282-9888]{Pieter van Dokkum}
\affiliation{Astronomy Department, Yale University, 52 Hillhouse Ave,
New Haven, CT 06511, USA}

\author[0000-0002-6442-6030]{David Weinberg}
\affiliation{The Department of Astronomy and Center of Cosmology and AstroParticle Physics, The Ohio State University, Columbus, OH 43210, USA}

\author[0000-0002-6442-6030]{Daniel R. Weisz}
\affiliation{Department of Astronomy, University of California, Berkeley, CA 94720, USA}


\begin{abstract}

We present the stellar metallicities and multi-element abundances (C, Mg, Si, Ca, Ti, Cr, and Fe) of 15 massive (log $M/M_\odot=10.2-11.2$) quiescent galaxies at $z=1-3$, derived from ultradeep \textit{JWST}-SUSPENSE spectra. Compared to quiescent galaxies at $z\sim0$, these galaxies exhibit a deficiency of 0.26$\pm0.04$ dex in [C/H], 0.16$\pm0.03$ dex in [Fe/H], and 0.07$\pm0.04$ dex in [Mg/H], implying rapid formation and quenching before significant enrichment from asymptotic giant branch stars and Type Ia supernovae. Additionally, we find that galaxies forming at higher redshift consistently show higher [Mg/Fe] and lower [Fe/H] and [Mg/H], regardless of their observed redshift. The evolution in [Fe/H] and [C/H] is therefore primarily driven by lower-redshift samples naturally including galaxies with longer star-formation timescales. In contrast, the lower [Mg/H] likely reflects earlier-forming galaxies expelling larger gas reservoirs during their quenching phase. Consequently, the mass-metallicity relation, primarily reflecting [Mg/H], is somewhat lower at $z=1-3$ compared to the lower redshift relation. Finally, we compare our results to standard stellar population modeling approaches employing solar abundance patterns and non-parametric star-formation histories (using \textsc{Prospector}). Our SSP-equivalent ages agree with the mass-weighted ages from \textsc{Prospector}, while the metallicities disagree significantly. Nonetheless, the metallicities better reflect [Fe/H] than total [Z/H]. We also find that star-formation timescales inferred from elemental abundances are significantly shorter than those from \textsc{Prospector}, and we discuss the resulting implications for the early formation of massive galaxies.

\end{abstract}


\section{Introduction} \label{sec:intro}
The chemical makeup of a galaxy is intimately linked to its past star formation, the amount of gas and stars it accretes, and the gas that is expelled through outflows. Consequently, the metal content of galaxies reflects fundamental physical processes that influence their evolution, such as star formation efficiency, feedback from massive stars and active galactic nuclei (AGN), and previous merger events. By characterizing how the metallicities of galaxies evolve across redshifts, we can obtain direct insights into the processes shaping galaxy growth throughout cosmic time.

In the nearby universe, galaxies display a tight correlation between their stellar mass and metallicity, known as the mass-metallicity relation \citep[MZR; e.g.,][]{lequeux_reprint_1979, tremonti_origin_2004, gallazzi_ages_2005, kirby_universal_2013}. For star-forming galaxies, the gas-phase MZR has been routinely studied out to $z\sim3$ mostly using strong-line indicators \citep[e.g.,][]{maiolino_amaze_2008, zahid_chemical_2013, sanders_mosdef_2020, sanders_mosdef_2021, papovich_clear_2022, shapley_jwstnirspec_2023} and has recently been confirmed up to $z\sim8$ with the advent of \textit{JWST} \citep[][]{langeroodi_evolution_2023, nakajima_jwst_2023, curti_jades_2024}. Quiescent galaxies, however, lack the strong emission lines needed for gas-phase measurements, so we instead rely on faint absorption lines originating from stellar atmospheres. At higher redshifts, measuring stellar metallicity becomes increasingly challenging as key absorption features shift to near-infrared (NIR) wavelengths, where ground-based spectroscopic observations are severely hindered by skylines. As a result, the MZR of quiescent galaxies has only been systematically studied up to $z\sim0.7$, showing little evolution since $z\sim0$ \citep{choi_assembly_2014,gallazzi_charting_2014, barone_lega-c_2022, beverage_carbon_2023}. 

Beyond $z\sim1$, the picture is much less clear. Currently, there are only a handful of stellar metallicity measurements of quiescent galaxies at $z=1-2.2$ based on absorption lines, mostly relying on methods such as spectral stacking \citep{onodera_ages_2015,carnall_stellar_2022}, observations of rare lensed massive galaxies \citep{jafariyazani_resolved_2020, man_exquisitely_2021, zhuang_glimpse_2023, akhshik_requiem-2d_2023, jafariyazani_chemical_2024}, low-resolution spectroscopy \citep{morishita_metal_2018, estrada-carpenter_clear_2019, akhshik_requiem-2d_2023}, or utilizing extreme integration times on the most efficient ground-based telescopes \citep{kriek_massive_2016, kriek_stellar_2019, kriek_heavy_2024, beverage_heavy_2024}. Despite the small samples at these redshifts, there is a growing consensus that massive quiescent galaxies at $z\gtrsim1$ exhibit significantly lower [Fe/H] compared to their present-day counterparts \citep[e.g.,][]{kriek_stellar_2019, zhuang_glimpse_2023, beverage_heavy_2024}. There are, however, a couple of conflicting findings, with two studies reporting exceptionally high metallicities comparable to the most metal-rich galaxies at $z\sim0$ \citep{lonoce_old_2015, jafariyazani_resolved_2020}. These contrasting results may not be surprising, given that the uncertainties in individual measurements remain high, and the methods vary significantly between studies \citep[e.g.,][]{lonoce_old_2015, onodera_ages_2015, kriek_massive_2016, estrada-carpenter_clear_2019, saracco_star_2023}.

Understanding the metallicities of distant quiescent galaxies is crucial, as it holds significant implications for the enrichment, star-formation timescales, quenching, and assembly of galaxies across cosmic time. Specifically, the multi-element abundance patterns provide direct insight into the star-formation histories (SFHs) of galaxies. Such insight is largely owed to the diverse enrichment timescales of the elements. $\alpha$ elements (e.g., O and Mg) are instantaneously released by core-collapse supernovae (CC SNe). C and N are released approximately equally by CC SNe and by the winds of low-mass asymptotic giant branch (AGB) stars which only contribute after a characteristic delay-time of $\sim250$ Myr \citep[e.g.][]{cristallo_evolution_2011, cristallo_evolution_2015, maiolino_re_2019, johnson_empirical_2023}. Fe-peak elements are enriched by both CC and the explosions of intermediate-mass stars (Type Ia SNe; Ia SNe), which occur only after a longer delay \citep[of $\sim1$~Gyr][]{maoz_supernova_2010}. 

To fully exploit chemical compositions for understanding the formation histories of massive galaxies, we need a larger sample of massive quiescent galaxies at $z\gtrsim1$ with deep spectra covering multiple absorption features. Such observations are now finally possible with \textit{JWST}. To that end, we have conducted the \textit{JWST}-SUSPENSE program, an ultradeep rest-frame optical spectroscopic NIRSpec/MSA survey of 20 massive quiescent galaxies at $z=1-3$ \citep[][]{slob_jwst-suspense_2024}. 

In this paper, we present the metallicities and multi-element abundances of distant quiescent galaxies at $z=1-3$ from the \textit{JWST}-SUSPENSE survey. In Section~\ref{sec:analysis} we describe the observations and elemental abundance analysis, in Section~\ref{sec:results} we present the abundance results, in Section~\ref{sec:mzr} we present the MZR at $z=1-3$, and in Section~\ref{sec:discuss} we discuss the implications of the results on star-formation timescales, the assembly of massive galaxies, and star-formation quenching. In Section~\ref{sec:discuss} we also compare the full-spectrum modeling results to those from \textsc{Prospector} and discuss the implications of this comparison on the early formation of massive galaxies. In Section~\ref{sec:summary}, we present a summary. Throughout this Paper, we assume a \citet{kroupa_variation_2001} IMF, solar abundances from \citet{asplund_chemical_2009}, and a flat $\Lambda$CDM cosmology with $\Omega_m=0.3$ and $H_0=70~\rm{km\,s^{-1}\,Mpc^{-1}}$.

\begin{figure*}
    \centering
    \includegraphics[width=\textwidth]{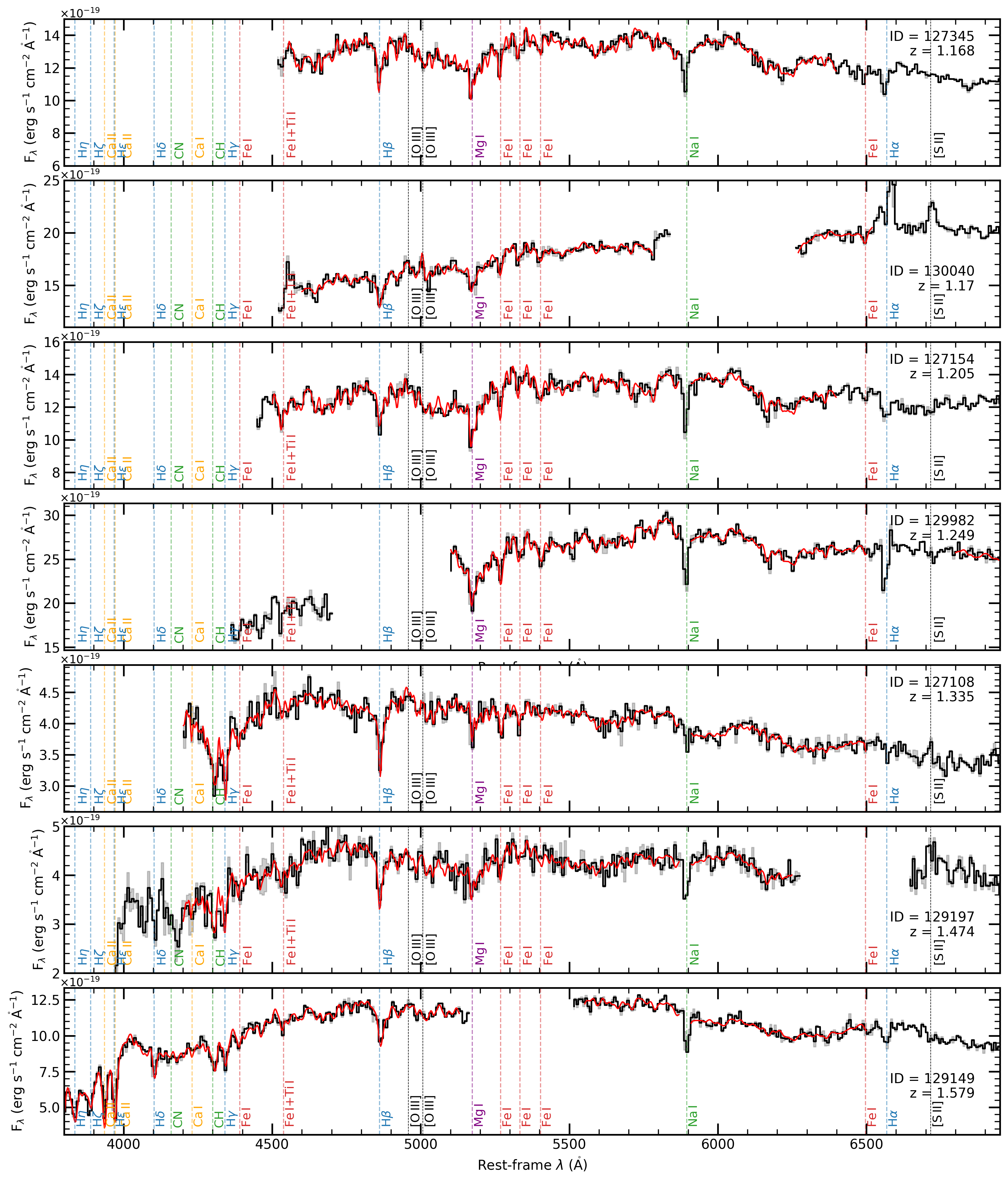}
    \caption{\textit{JWST}-SUSPENSE spectra of quiescent galaxies at $z=1-3$ (black) and corresponding 1$\sigma$ uncertainties (gray). The best-fit stellar population models are shown in red. The wavelength array has been corrected for redshift and is shown in the rest frame. The spectra have been median binned by 3 pixels (8 \AA~in the rest frame). }
    \label{fig:app_spec1}
\end{figure*}

\begin{figure*}
    \ContinuedFloat
    \captionsetup{list=off}
    \centering
    \includegraphics[width=\textwidth]{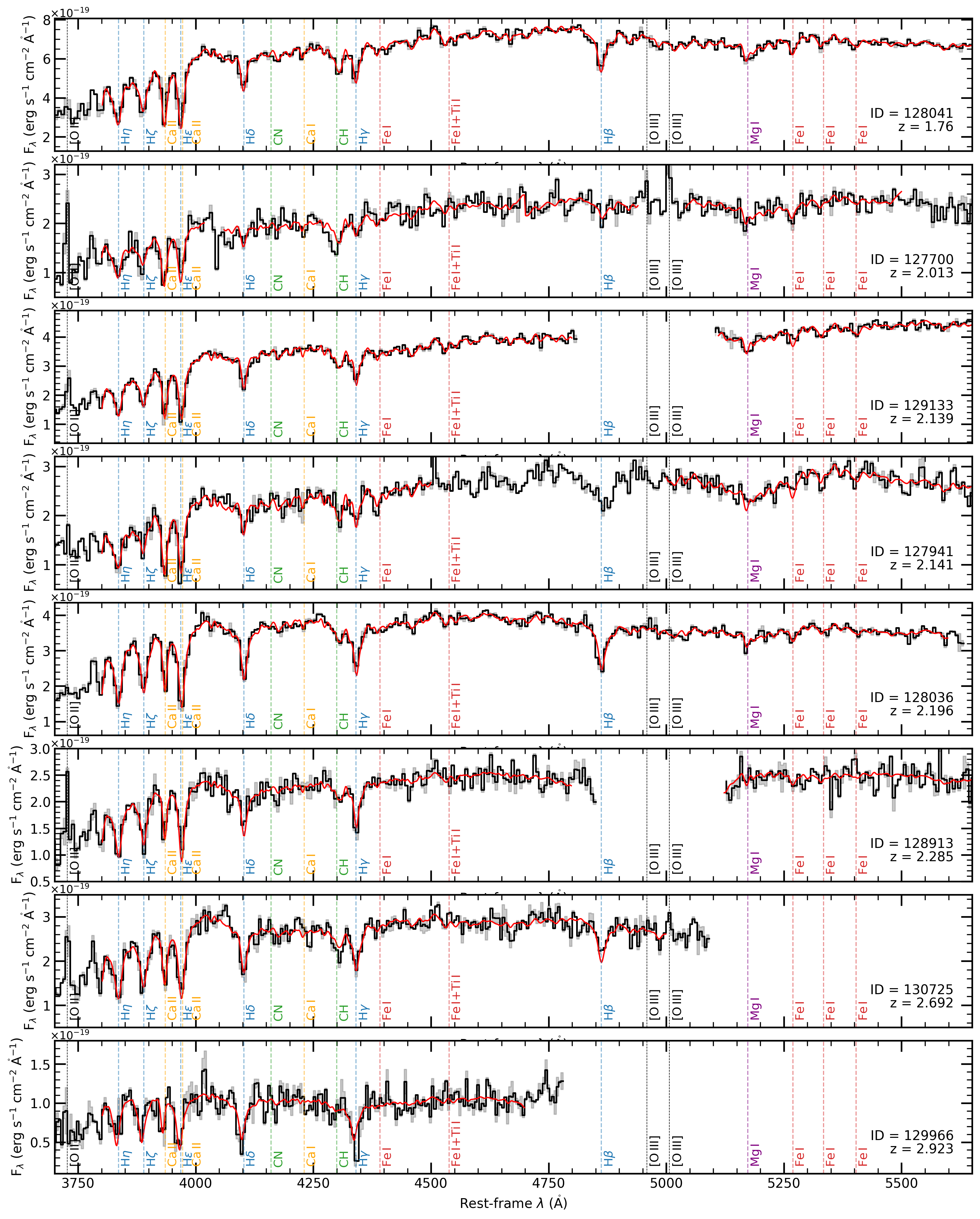}
    \caption{Continued}
\end{figure*}

\begin{deluxetable*}{cccccccccccc}
\tabletypesize{\footnotesize}\tablecolumns{8}
\tablewidth{0pt}
\label{table:alf-results}
\tablecaption{\label{table:alf-results}JWST-SUSPENSE stellar population parameters}
\tablehead{\colhead{ID}\vspace{-0.2cm} & \colhead{$z_\mathrm{spec}^{a}$} & \colhead{$\log M^{a}$} & \colhead{$\sigma$} & \colhead{Age} & \colhead{[Fe/H]} & \colhead{[Mg/H]}\ & \colhead{[C/H]}\ & \colhead{[Ca/H]}\ & \colhead{[Ti/H]}\ & \colhead{[Cr/H]}\ & \colhead{[Si/H]}\\ 
\colhead{} & \colhead{} & \colhead{($M_\odot$)} & \colhead{(km\,s$^{-1}$)} & \colhead{(Gyr)} & \colhead{} & \colhead{} & \colhead{} & \colhead{} & \colhead{} & \colhead{} & \colhead{}}
\startdata
127345 & 1.168 & 10.7 & $166^{+11}_{-12}$ & $2.6^{+0.4}_{-0.3}$ & $-0.17^{+0.08}_{-0.08}$ & $0.15^{+0.10}_{-0.10}$ & $-0.03^{+0.09}_{-0.10}$ & $0.14^{+0.09}_{-0.08}$ & -- & $-0.19^{+0.12}_{-0.12}$ & $-0.10^{+0.12}_{-0.13}$\\ 
130040 & 1.170 & 11.1 & $259^{+18}_{-17}$ & $2.0^{+0.3}_{-0.3}$ & $-0.18^{+0.08}_{-0.07}$ & $0.23^{+0.13}_{-0.13}$ & $0.07^{+0.13}_{-0.15}$ & $-0.04^{+0.11}_{-0.11}$ & $0.38^{+0.19}_{-0.23}$ & $0.04^{+0.17}_{-0.16}$ & $-0.07^{+0.14}_{-0.13}$\\ 
127154 & 1.205 & 10.8 & $206^{+11}_{-11}$ & $2.2^{+0.2}_{-0.2}$ & $0.17^{+0.08}_{-0.08}$ & $0.55^{+0.10}_{-0.10}$ & $0.30^{+0.07}_{-0.09}$ & -- & $0.47^{+0.14}_{-0.16}$ & $0.06^{+0.13}_{-0.13}$ & $0.39^{+0.10}_{-0.11}$\\ 
129982 & 1.249 & 11.2 & $260^{+19}_{-18}$ & $4.2^{+1.3}_{-1.2}$ & $-0.10^{+0.10}_{-0.10}$ & $0.24^{+0.14}_{-0.14}$ & $-0.05^{+0.16}_{-0.16}$ & $0.04^{+0.12}_{-0.14}$ & $0.06^{+0.21}_{-0.21}$ & $0.19^{+0.16}_{-0.16}$ & --\\ 
127108 & 1.335 & 10.3 & $177^{+17}_{-16}$ & $3.4^{+0.6}_{-0.5}$ & $-0.55^{+0.10}_{-0.10}$ & $-0.24^{+0.12}_{-0.13}$ & $-0.77^{+0.14}_{-0.14}$ & $-0.61^{+0.19}_{-0.18}$ & -- & $-0.95^{+0.14}_{-0.13}$ & $-0.23^{+0.18}_{-0.23}$\\ 
129197 & 1.474 & 10.6 & $145^{+25}_{-23}$ & $2.1^{+0.5}_{-0.4}$ & -- & -- & -- & -- & -- & -- & --\\ 
129149 & 1.579 & 11.0 & $290^{+12}_{-12}$ & $1.9^{+0.2}_{-0.2}$ & $-0.12^{+0.08}_{-0.09}$ & -- & $0.01^{+0.10}_{-0.11}$ & $-0.19^{+0.07}_{-0.07}$ & $0.23^{+0.15}_{-0.18}$ & $0.06^{+0.16}_{-0.15}$ & $-0.01^{+0.16}_{-0.18}$\\ 
128041 & 1.760 & 10.7 & $240^{+8}_{-8}$ & $1.6^{+0.2}_{-0.1}$ & $-0.17^{+0.09}_{-0.09}$ & $0.27^{+0.09}_{-0.09}$ & $-0.18^{+0.10}_{-0.11}$ & $-0.12^{+0.06}_{-0.06}$ & $0.17^{+0.12}_{-0.13}$ & $-0.27^{+0.12}_{-0.12}$ & $-0.21^{+0.12}_{-0.14}$\\ 
127700 & 2.013 & 10.8 & $236^{+36}_{-30}$ & $3.2^{+0.2}_{-0.3}$ & $-0.15^{+0.10}_{-0.12}$ & -- & $-0.24^{+0.14}_{-0.15}$ & $-0.18^{+0.11}_{-0.11}$ & -- & -- & --\\ 
129133 & 2.139 & 11.1 & $225^{+10}_{-10}$ & $1.2^{+0.1}_{-0.1}$ & $-0.17^{+0.08}_{-0.08}$ & $0.33^{+0.08}_{-0.09}$ & $0.01^{+0.09}_{-0.09}$ & $0.22^{+0.06}_{-0.06}$ & -- & $-0.14^{+0.15}_{-0.14}$ & $0.07^{+0.14}_{-0.15}$\\ 
127941 & 2.141 & 11.0 & $197^{+20}_{-19}$ & $2.0^{+0.4}_{-0.3}$ & $-0.43^{+0.17}_{-0.17}$ & $-0.33^{+0.22}_{-0.19}$ & $-0.48^{+0.21}_{-0.21}$ & $0.18^{+0.10}_{-0.10}$ & -- & $0.07^{+0.21}_{-0.22}$ & $0.01^{+0.18}_{-0.22}$\\ 
128036 & 2.196 & 11.0 & $220^{+9}_{-9}$ & $1.1^{+0.1}_{-0.1}$ & $-0.26^{+0.09}_{-0.10}$ & $0.06^{+0.12}_{-0.12}$ & $-0.27^{+0.11}_{-0.12}$ & $0.15^{+0.05}_{-0.05}$ & -- & $-0.45^{+0.18}_{-0.18}$ & $-0.25^{+0.18}_{-0.21}$\\ 
128913 & 2.285 & 10.8 & $187^{+28}_{-27}$ & $2.3^{+0.5}_{-0.5}$ & -- & -- & -- & -- & -- & -- & --\\ 
130725 & 2.692 & 11.2 & $248^{+22}_{-22}$ & $1.1^{+0.2}_{-0.2}$ & $-0.35^{+0.21}_{-0.23}$ & -- & -- & $-0.09^{+0.12}_{-0.13}$ & -- & -- & --\\ 
129966$^*$ & 2.923 & 10.9 & $207^{+44}_{-45}$ & $0.6^{+0.1}_{-0.1}$ & -- & -- & -- & -- & -- & -- & --\\ 
\enddata
\tablenotetext{a}{Presented in \citet{slob_jwst-suspense_2024}}
\vspace{-0.25cm}
\tablenotetext{*}{Removed from analysis because younger than the stellar population model grid ($<1$~Gyr)}
\vspace{-0.5cm}
\end{deluxetable*}

\begin{figure*}
    \centering
    \includegraphics[width=1\textwidth]{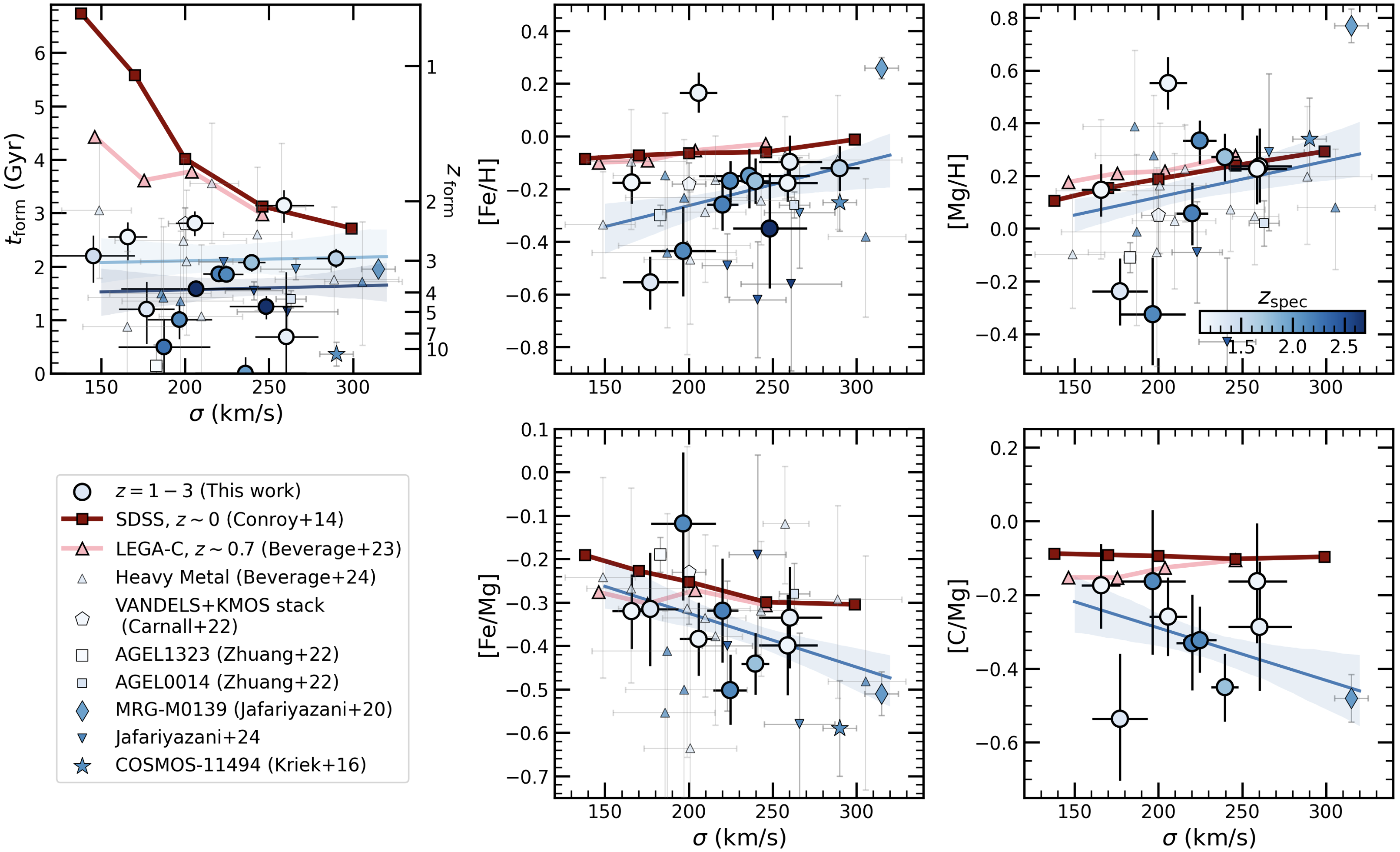}
    \caption{The formation time ($t_{\rm form}$), [Fe/H], [Mg/H], [Fe/Mg], and [C/Mg] as a function of velocity dispersion for the $z=1-3$ \textit{JWST}-SUSPENSE quiescent galaxy sample (circles) and for other various measurements at similar redshifts from the literature. Each point is colored by its spectroscopic redshifts. To guide the eye, we include the best-fit relations and corresponding confidence intervals in each panel, fit to the $z=1-3$ data point. In the $t_{\rm form}$ panel, we instead present two relations, one for $z<1.5$ and the other for $z>1.5$. For comparison, we also include the abundances of stacks of quiescent galaxies at $z\sim0$ from SDSS \citep{conroy_early-type_2014,beverage_carbon_2023} and at $z\sim0.7$ from LEGA-C \citep{beverage_carbon_2023}. At constant $\sigma$, the SUSPENSE galaxies form earlier and are more deficient in Fe and C than the $z<0.7$ galaxies.}
    \label{fig:abundance_panels}
\end{figure*}

\begin{figure*}
    \centering
    \includegraphics[width=\textwidth]{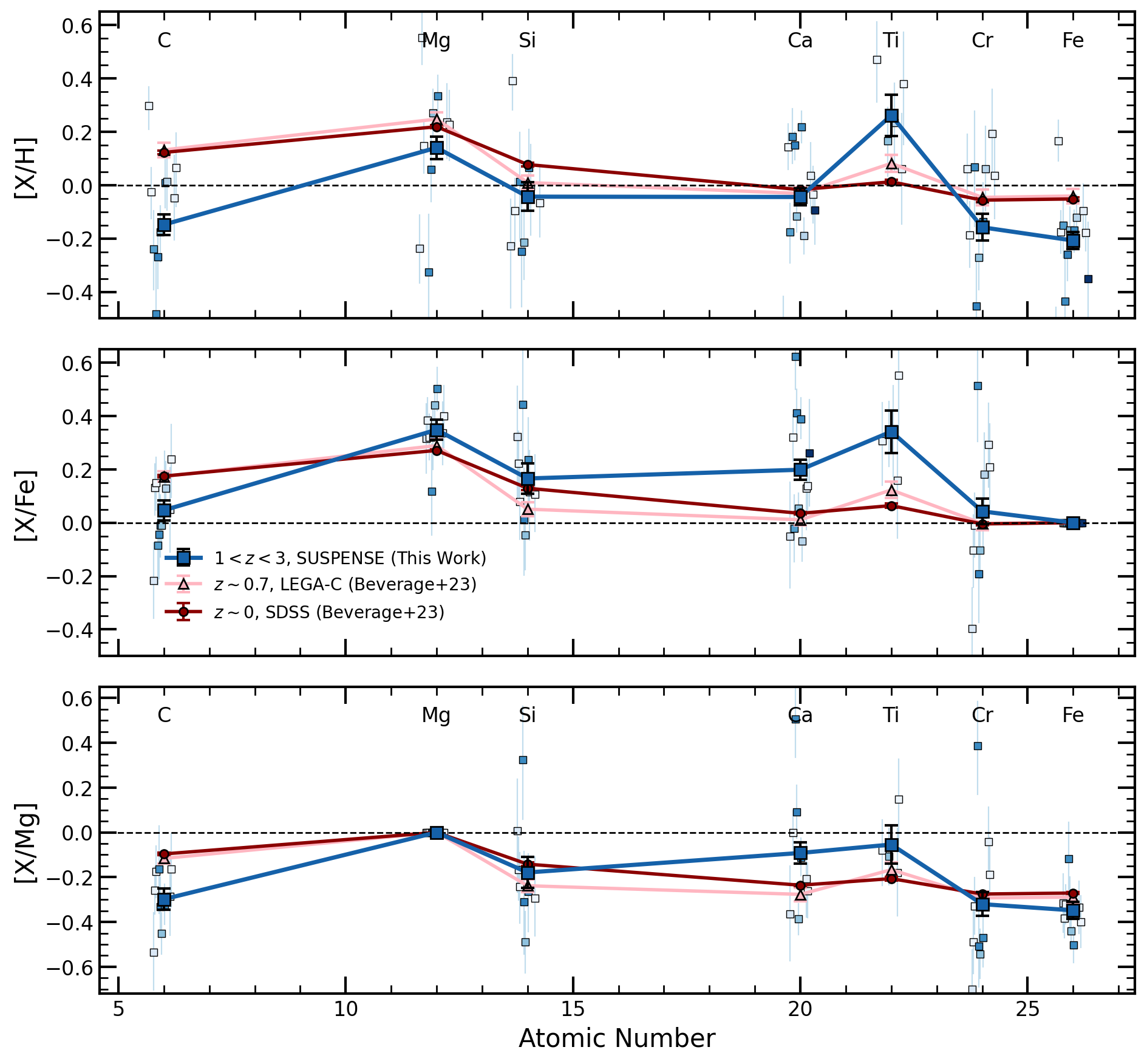}
    \caption{The abundance patterns of the $z=1-3$ quiescent galaxies (blue squares). We show the absolute abundances, [X/H], in the top panel, followed by the abundance ratios [X/Fe] (middle) and [X/Mg] (bottom), with the solar values marked with a dashed black line. The thick solid blue line with square markers shows the median abundance pattern of the $z=1-3$ quiescent galaxies. For comparison, we include the average abundance patterns of $z\sim0$ (red solid line with circle markers) and $z\sim0.7$ (pink solid line with triangle markers) galaxies with the \textit{same} velocity dispersions ($175-275~\mathrm{km\,s^{-1}}$). The $z=1-3$ galaxies are significantly deficient in [C/H] and [Fe/H] compared to lower redshift samples. }
    \label{fig:abundance_pattern}
\end{figure*}

\vspace{-0.8cm}

\section{Observations \& Analysis} \label{sec:analysis}

All galaxies in this study are drawn from the \textit{JWST}-SUSPENSE Program (ID:2110) which obtained ultra-deep (16.4 hr) NIRSpec-MSA/G140M-F100LP observations of a sample of 20 massive quiescent galaxies at $z=1-3$ \citep{slob_jwst-suspense_2024}. The spectroscopic observations were reduced using a modified version of the \textit{JWST} Science Calibration Pipeline \citep{bushouse_2023_10022973} v1.12.5, and version 1183 of the Calibration Reference Data System (CRDS) \citep[see][for details]{slob_jwst-suspense_2024}. Primary targets were initially identified using the UltraVISTA K-band selected DR3 catalog \citep[][]{muzzin_public_2013} and selected to be quiescent using the $UVJ$ criterion from \citet{muzzin_evolution_2013}. The targets represent the general population of quiescent galaxies at $z=1-3$, encompassing the full quiescent range in $UVJ$ space at these redshifts, and are all confirmed to have quiescent stellar populations \citep[see][]{slob_jwst-suspense_2024}{}{}.

The observations cover wavelengths from $0.97$ to $1.84,\mu$m, corresponding to a typical rest-frame range of approximately $3700$ to $7000$ \AA. Most objects have coverage of the Mgb line, several prominent Fe features, and multiple Balmer lines. The stellar masses and star-formation rates were derived by \citet{slob_jwst-suspense_2024} by fitting the spectra and photometry simultaneously with \textsc{Prospector} \citep{leja_how_2019, johnson_stellar_2021} assuming a \citet{chabrier_galactic_2003} IMF. They consider two non-parametric SFH models; a fixed-bin model \citep{leja_how_2019} and a post-starburst model \citep{suess_squiggle_2022}. For more details on target selection, observing strategy, data reduction, redshift determination, and stellar population properties we refer to \citet{slob_jwst-suspense_2024}.

We measure the individual elemental abundances and stellar population ages using a custom full-spectrum fitting code \texttt{alf$\alpha$} \citep[][]{aliza_beverage_alizabeveragealfalpha_2024}\footnote{\href{https://github.com/alizabeverage/alfalpha}{https://github.com/alizabeverage/alfalpha}} based on the \texttt{alf} fitting code presented in \citet{conroy_metal-rich_2018}. The \citet{conroy_metal-rich_2018} models were developed to measure the elemental abundance patterns of old ($\gtrsim$\,1\;Gyr) stellar populations. They combine metallicity-dependent MIST isochrones \citep{choi_mesa_2016}, empirical MILES and IRTF spectral libraries \citep{sanchez-blazquez_medium-resolution_2006,villaume_extended_2017}, and synthetic metallicity- and age-dependent elemental response spectra for 19 elements. 

The models assume a single burst of star formation and a \citet{kroupa_variation_2001} IMF. We fit for a total of 20 free parameters; velocity offset, velocity dispersion, single SSP-equivalent stellar population age, isochrone metallicity, 10 individual elemental abundances (Fe, C, N, Mg, Na, Si, Ca, Ti, Cr), Balmer emission line flux, the emission line velocity and broadening, a shift in the effective temperature of the fiducial isochrones ($T_{\rm eff}$), and an instrumental jitter term to account for under/overestimation of the uncertainties. The abundance ratios of the other nine elements are fixed to solar ([X/Fe] = 0) following results of the mock-recovery test presented in \citet{beverage_carbon_2023}. We use the \texttt{dynesty} dynamic nested sampling package \citep{speagle_dynesty_2020} to sample the posterior distributions of the 20 parameters.

Before fitting, we smooth the models to the instrumental resolution of the observations. We derive the instrumental resolution as a function of wavelength in the raw spectral frames using \texttt{msafit} \citep[][]{de_graaff_ionised_2024}. We provide \texttt{msafit} with the morphology of each galaxy, its position and angle with respect to the MSA slit, and a spectrum filled with equally spaced delta functions that simulate idealized emission lines. The software then simulates a 2D spectrum for each object. We process these 2D spectra using the \textit{JWST} NIRSpec reduction pipeline in the same manner as the actual data. Finally, we measure the spectral resolution by analyzing the broadening of the simulated emission lines. We find that the wavelength dependence of the instrumental resolution is consistent with the online pre-launch estimates from JDox, but typically a factor of 1.3 better, corresponding to a resolution of $R\sim1300$. For each galaxy, we derive this factor and assume the corrected JDox curve when fitting. See \citet{slob_jwst-suspense_2024} for more details on this procedure.

During the fitting procedure, the spectral continuum is removed from the observations by fitting an $n=7$ Chebyshev polynomial to the ratio of the data to the model. We note that the results are robust to decisions regarding the order of the polynomial. Where available, we fit the wavelength regions $3800-4800$\,\AA, $4800-5800$\,\AA, $5800-6400$\,\AA, and $8000-8600$\,\AA\ (each with their own normalized continuum). We mask the NaD absorption feature, the [OIII] lines where present, and the H$\alpha$+[NII] complex. We also exclude wavelengths $6400-8000$\,\AA\ due to the dominant TiO absorption in the spectrum, as these broad features are typically over-fitted by the continuum polynomial. 

After fitting all 20 galaxies, we visually inspect the best-fit model, the normalizing polynomial, and the corresponding corner plots. We remove five galaxies; two because they have low S/N ($<$15 per rest-frame \AA; 130183, 130934), one because it is lacking wavelength coverage of key absorption features (130208), and two because of strong emission lines likely associated with AGN (130647, 128452). We also re-fit 127941, masking the H$\beta$ region, which was poorly fit due to spectral contamination. Object 127108 is also re-fit, masking the region $>5500\,$\AA\ due to the lack of clear features. Finally, for each object, we inspect the posteriors of all 10 fitted elements and determine which elements can be constrained by requiring their posterior to be Gaussian and not run against the prior limits ($-$0.5 and 0.5 dex). Figure~\ref{fig:app_spec1} shows the quiescent SUSPENSE spectra at $z=1-3$ and corresponding best-fit \texttt{alf$\alpha$} models. We present the stellar population properties and elemental abundances [X/H] in Table~\ref{table:alf-results}. The elements flagged as ``poorly constrained'' during the visual inspection are omitted from Table~\ref{table:alf-results}. We also omit the abundances of 129966 because the best-fit age is below the lowest age in the stellar population models (1~Gyr). Tables of the elemental abundance ratios [X/Fe] and [X/Mg] are provided in Appendix~\ref{app:table}.


\section{The multi-elemental abundances of distant quiescent galaxies}
\label{sec:results}

In this section, we present the multi-element abundances of the massive quiescent galaxies at $z=1-3$ from the \textit{JWST}-SUSPENSE program. For each SUSPENSE galaxy, we only consider the elements that have well-behaved normal posterior distributions. Among the ten fitted elements, C, Mg, and Fe have the most galaxies with well-constrained measurements, likely due to their prominent absorption features, such as CH (G4300), Mgb, Fe5270, Fe5335, and Fe5406. 

In Figure~\ref{fig:abundance_panels}, we present the formation time ($t_\mathrm{form}$), [Fe/H], [Mg/H], [Fe/Mg], and [C/Mg] as a function of line-of-sight velocity dispersion (circles). Here, we adopt Mg as the reference element instead of Fe (i.e., [X/Mg]) because it is mostly produced by CC SNe and thus a simpler tracer of chemical enrichment \citep[see e.g.,][]{weinberg_chemical_2019}. The formation time is calculated using the stellar ages from Table~\ref{table:alf-results}, corrected for the age of the universe at the observed redshift. The points are colored by their spectroscopic redshifts. We also include the results of stacked quiescent galaxies at $z\sim0$ from SDSS (red squares) and $z\sim0.7$ from LEGA-C \citep[pink triangles][]{beverage_carbon_2023}. The stacked SDSS spectra and corresponding elemental abundance results were first presented in \citet{conroy_early-type_2014} and later re-fit by \citet{beverage_carbon_2023} to reflect updates in the stellar population models \citep[see][]{conroy_metal-rich_2018}. We note that the SDSS fiber and the LEGA-C and SUSPENSE slits all cover similar physical radii (3 -- 4 kpc). However, since galaxies were smaller in the past, the SUSPENSE and LEGA-C slits cover 1 -- 1.2 $R_e$, while SDSS covers only 0.4 -- 0.8 $R_e$.

In Figure~\ref{fig:abundance_panels}, we also include all available elemental abundance measurements in the literature at $z=1-3$. We only include measurements that use the same full-spectrum fitting method, and thus are comparable to the SUSPENSE results. We do not include grism or prism results because they rely heavily on the shape of the continuum and are therefore highly susceptible to fitting degeneracies. We also exclude abundances derived from Lick indices because, at these redshifts, the individual absorption features are faint and easily contaminated by NIR skylines. We color all points by their spectroscopic redshifts.

In each panel of Figure~\ref{fig:abundance_panels}, we present the results from a linear structural regression fit applied to \textit{all} available points within the redshift range $z=1-3$. The linear model assumes intrinsic scatter around the best-fit line, which is described by a third parameter alongside the slope and intercept. Using a Bayesian approach, we determine the best-fit line and its associated shaded confidence intervals by sampling the posterior distributions with MCMC. We fit two relations to the formation times, one at $z<1.7$ (light blue) and another at $z>1.7$ (dark blue). During the fitting process, we set a lower limit of 0.1~dex on the uncertainties of the elemental abundance, to prevent single high S/N measurements from dominating the fit and to account for possible systematic uncertainties. 

In the top left panel of Figure~\ref{fig:abundance_panels}, we find that the formation redshifts of the SUSPENSE galaxies range from $z_{\rm form} = 1.5$ to $9$. No clear trends are observed between $\sigma$ and formation time at $z \gtrsim 1$. However, there is a clear trend with observed redshift, where typical galaxies at $z \gtrsim 1$ formed earlier than those at lower redshifts, as expected. This redshift trend is also evident within the $z \gtrsim 1$ sample, with the best-fit relation for $z > 1.7$ being lower than that for $z < 1.7$. A similar trend has been shown by \citet{beverage_carbon_2023} and will be discussed in detail in Section~\ref{sec:assembly}

In the next panel, we show that quiescent galaxies at $z=1-3$ have [Fe/H] ranging from $-$0.6 to 0.3, with typical values of $-$0.22. We find a trend between $\sigma$ and [Fe/H], with a statistically significant slope (2$\sigma$ certainty). The normalization of these measurements are 0.16$\pm0.03$~dex lower than what is found at $z\sim0$ and $z\sim0.7$. A 1D Kolmogorov-Smirnov (KS) test comparing the Fe abundances of galaxies at  $z \sim 0$  and  $z \sim 2$  yields significant p-values ($<0.01$), allowing us to reject the null hypothesis and conclude that the Fe abundances of these two populations are drawn from different distributions. Thus, we confirm earlier results that quiescent galaxies at $z\sim2$ are Fe-deficient \citep{kriek_massive_2016, kriek_stellar_2019, morishita_metal_2018, zhuang_glimpse_2023, beverage_heavy_2024, jafariyazani_chemical_2024}. In the next panel, we find that [Mg/H] varies from $-$0.3 to 0.6, with a typical value of [Mg/H]$=0.15$. The slope of the $\sigma-$[Mg/H] relation agrees with the results at $z\sim0$ and $z\sim0.7$ and the normalization is only slightly lower (0.07$\pm0.04$~dex). 

\begin{figure*}
    \centering
    \includegraphics[width=\textwidth]{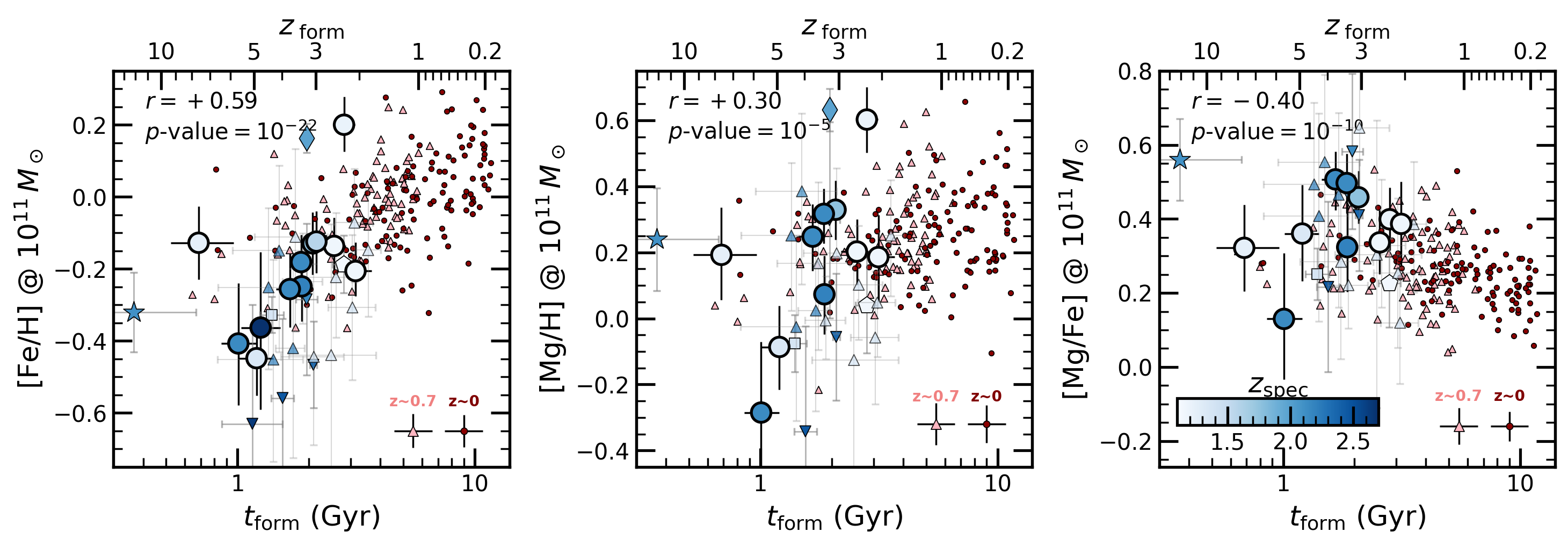}
    \caption{[Fe/H], [Mg/H], and [Mg/Fe] as a function of galaxy formation time for massive quiescent galaxies at $z=0-3$. Data points for the redshift range $z=1-3$ are color-coded based on their spectroscopic redshift, and their shapes follow the same legend as Figure~\ref{fig:abundance_panels}. The red circles [pink triangles] represent individual quiescent galaxies from SDSS [LEGA-C] \citep{zhuang_glimpse_2023, beverage_carbon_2023}. Typical uncertainties for the SDSS and LEGA-C points are shown in the bottom right of each panel. We remove the first-order dependence on stellar mass by adjusting all measurements to reflect the abundances at $M_*=10^{11}\;M_\odot$ (refer to the text for details). We calculate the Pearson correlation coefficients, including all galaxies at all redshifts, and list the corresponding correlations (r) and $p$-values in the top left corner of each panel. Regardless of observed redshift, galaxies that form earlier have lower [Fe/H], slightly lower [Mg/H], and higher [Mg/Fe], consistent with more rapid formation and efficient quenching at earlier times.}
    \label{fig:formation_time-label}
\end{figure*}

\begin{figure*}
    \centering
    \includegraphics[width=0.8\textwidth]{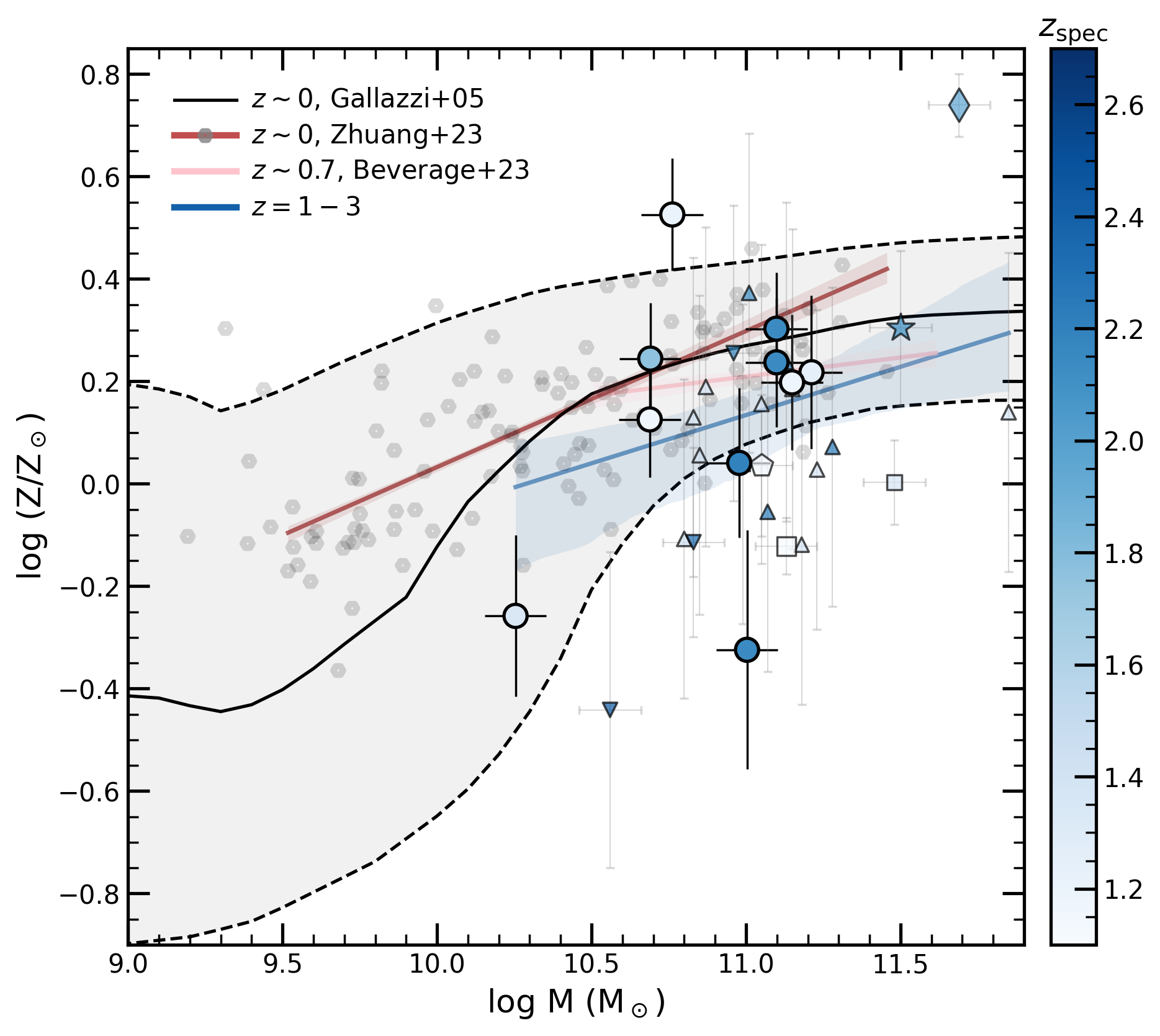}
    \caption{Stellar metallicity as a function of stellar mass for quiescent galaxies at $z=1-3$ in the \textit{JWST}-SUSPENSE program (circles) and from various other studies (following the same legend as Figure~\ref{fig:abundance_panels}). All points are colored by their spectroscopic redshifts. The blue line shows the best-fit stellar mass-metallicity relation at $z=1-3$, measured from all available measurements at $z=1-3$. We include the stellar MZR at $z\sim0.7$ from \citet[][black dashed line]{beverage_carbon_2023} and $z\sim0$ from \citet[][black line and gray shaded region]{gallazzi_ages_2005}. We also show individual values from the SDSS sample presented in \citet[][gray hexagons]{zhuang_glimpse_2023}. Galaxies $z=1-3$ on average have $-0.17\pm0.09$\;dex lower metallicities than those at $z\sim0$ and $z\sim0.7$.}
    \label{fig:mzr}
\end{figure*}

In the bottom left panel of Figure~\ref{fig:abundance_panels} we find [Fe/Mg] ranging from $-0.50$ to $-$0.10, with a typical value of $-$0.35. The best-fit $\sigma$-[Fe/Mg] relation has a negative slope, in agreement with lower redshift results, however, the relation is offset to lower [Fe/Mg] by 0.10$\pm0.04$ dex. A 1D KS test of [Fe/Mg] show that the $z\sim0$ and $z\sim2$ abundances are drawn from different distributions, with a mildly significant p-value ($<0.05$). This low [Fe/Mg] is consistent with the other results at similar redshifts in this figure \citep{kriek_massive_2016,beverage_heavy_2024,jafariyazani_resolved_2020,jafariyazani_chemical_2024}.

In the bottom right panel, we show [C/Mg]. We highlight that these are the first measurements of C abundances for a sample of distant quiescent galaxies and that only one other [C/Mg] measurement at these redshifts is available in the literature \citep{jafariyazani_resolved_2020}. We find [C/Mg] ranges from $-0.55$ to $-0.05$, with a typical value of $-0.3$, in good agreement with \citet{jafariyazani_resolved_2020}. The best-fit trend with $\sigma$ is still highly uncertain given that there are only 10 C measurements at this redshift but it is clear that the SUSPENSE galaxies have $\sim0.2$ dex lower [C/Mg] than the $z\sim0$ and $z\sim0.7$ samples (3$\sigma$ certainty). Given that the average [Mg/H] is similar (to within 0.07~dex) across the redshift samples, the low [C/Mg] and [Fe/Mg] are caused by deficiencies in C and Fe. 

Next, we present the elemental abundance patterns of the SUSPENSE galaxies in Figure~\ref{fig:abundance_pattern}. We show the average absolute abundances [X/H] (top panel) the abundance ratios with respect to Fe [X/Fe], (middle panel), and the abundance ratios with respect to Mg [X/Mg] (bottom panel). We only include elements for which we have at least five constrained individual measurements; namely C, Mg, Si, Ca, Ti, and Cr, and Fe. The small blue squares represent the individual abundances from SUSPENSE, while the large squares depict the mean of these measurements. The black error bars are calculated by perturbing the abundances of each data point based on their uncertainties and then determining the standard deviation of the resulting mean abundance values. Figure~\ref{fig:abundance_pattern} also includes the abundance patterns of $z\sim0$ (red circles) and $z\sim0.7$ quiescent galaxies (pink triangles) from \citet{beverage_carbon_2023}. These $z\sim0$ [$z\sim0.7$] results represent the average of thousands [hundreds] of quiescent galaxies from the \textit{same} velocity dispersion range as the SUSPENSE galaxies (200 -- 300 km\,$^{-1}$). 

Consistent with the findings of Figure~\ref{fig:abundance_panels}, massive quiescent galaxies at all three redshift intervals have non-solar abundance patterns, being deficient with respect to Mg. These trends are further evident when examining the individual elements in greater detail. In the top panel of Figure~\ref{fig:abundance_pattern}, we find that distant quiescent galaxies in SUSPENSE have significantly lower [C/H] and [Fe/H] ($>3\sigma$ certainty) and slightly lower [Mg/H] (1.5$\sigma$) than the $z<1$ galaxies. Specifically, [C/H], [Fe/H], and [Mg/H] are lower by 0.25, 0.16, and 0.07 dex, respectively, than at $z \sim 0$. Therefore, the [Mg/Fe] enhancement observed in the central panel is primarily driven by a deficiency in Fe. Moving on to the other elements, we find that Ca has a similar behavior as Mg, with a mostly constant [Ca/H] across the three redshift regimes, but with enhanced [Ca/Fe] and [Ca/Mg] at higher redshift. Ti is unique because the absolute abundance [Ti/H] and abundance ratios [Ti/Fe] and [Ti/Mg] increase with increasing redshift. We highlight that high Ti abundances present a long-standing problem in chemical evolution modeling \citep[e.g.,][]{kobayashi_origin_2020}. [Cr/H] on the other hand behaves more like Fe and C, with [Cr/H] being lower at $z=1-3$. This result is unsurprising given that Cr is typically considered an Fe-peak element. Si shows no significant evolution in [X/H], [X/Fe], or [X/Mg]. Thus, Si behaves more similarly to Mg and Ca. However, the abundance uncertainties on Ti, Si, and Cr are quite large. We note that these abundance results are roughly consistent with the only other existing abundance pattern at $z\gtrsim0.7$, observed in a lensed quiescent galaxy by \citet{jafariyazani_resolved_2020}. In Section~\ref{sec:timescale}, we explore the implications of these abundance patterns in the context of chemical enrichment histories.

Finally, in Figure~\ref{fig:formation_time-label}, we show [Fe/H], [Mg/H], and [Mg/Fe] as a function of formation time, as in \citet{zhuang_glimpse_2023}. In addition to the SUSPENSE sample (blue circles, colored by spectroscopic redshift), we include individual results from massive quiescent galaxies in the SDSS survey from \citet[][red circles]{zhuang_glimpse_2023} and LEGA-C from \citet[][pink triangles]{beverage_carbon_2023}. In each panel of Figure~\ref{fig:formation_time-label}, we remove the first-order dependence on stellar mass by subtracting the SDSS $M_*-$abundance relations from \citet{zhuang_glimpse_2023} from all galaxies and then scaling them to the value of the $M_*-$abundance relations at $M_* = 10^{11}\;M_\odot$. 
We utilize the Pearson correlation coefficient to assess the degree of correlation between the parameters in Figure~\ref{fig:formation_time-label}. We include all points in Figure~\ref{fig:formation_time-label} (at all redshifts) when computing the degree of correlation. The resulting Pearson coefficients and corresponding $p$-values are listed in the top left of each panel. Correlations with $p$-values $p<0.01$ are considered significant. 

At fixed stellar mass, the formation time and [Fe/H] have a significant positive correlation ($r=0.59$), with galaxies forming at earlier times having lower Fe enrichment. [Mg/H] shows a moderate positive correlation ($r=0.30$), with galaxies forming earlier having slightly lower Mg enrichment. Finally, there is a negative correlation ($r=-0.40$) between [Mg/Fe] and formation time, indicating galaxies that form earlier have slightly higher [Mg/Fe]. These correlations are primarily driven by the large quantity and dynamic range in $t_\mathrm{form}$ of $z\sim0$ abundance measurements and thus agree with the results presented in \citet{zhuang_glimpse_2023}. Therefore, it is striking that the correlations found at lower redshifts are mostly consistent with the $z\sim2$ data, in the sense that the oldest galaxies at $z\sim0$ have similar abundance patterns as those at $z\sim2$. In Section~\ref{sec:assembly}, we discuss the implications of these results on the assembly of massive galaxies.

\section{The Stellar Mass-Metallicity Relation at \texorpdfstring{$\lowercase{z}\sim2$}{z~2}}
\label{sec:mzr}

In this section, we present the stellar mass-metallicity relation of quiescent galaxies at $z=1-3$. In Figure~\ref{fig:mzr}, we show the stellar metallicities $\log\left(Z/Z_\odot\right)$ of the \textit{JWST}-SUSPENSE quiescent galaxies as a function of their stellar mass (circles). We compute these metallicities following the calibration $\log\left(Z/Z_\odot\right) =$~[Fe/H]$+0.94$[Mg/Fe] from \citet{thomas_stellar_2003}. As in Figure~\ref{fig:abundance_panels}, we include all existing measurements of stellar metallicities derived from full-spectrum fitting at $z=1-3$ derived using the same full-spectrum fitting method and underlying models. For comparison, we also include the $z\sim0$ MZR and corresponding 1$\sigma$ scatter from \citet[][black line and gray shaded region]{gallazzi_ages_2005}. Additionally, we show individual measurements for a subsample of these SDSS galaxies from \citet[][gray hexagons]{zhuang_glimpse_2023}. The \citet{gallazzi_ages_2005} relation was measured using a different method than \citet{zhuang_glimpse_2023}, with different stellar population models, and assuming a different solar abundance pattern. Therefore, we calibrate the \citet{gallazzi_ages_2005} MZR to match the normalization of the \citet{zhuang_glimpse_2023} measurements by applying a shift of +0.16 dex. 

We fit a mass-metallicity relation to all available $z=1-3$ measurements using the same structural linear regression applied to the abundances in Figure~\ref{fig:abundance_panels}, where we assume that the points have an intrinsic scatter around the best-fit line. Again, we set the lower limit on the error of the metallicity to 0.1 dex so that high S/N measurements do not dominate the fit. Uncertainties on all stellar mass estimates are assumed to be 0.1~dex, to properly account for systematic uncertainties involved in the stellar population modeling \citep[see][for motivation of mass uncertainties]{kriek_massive_2016}. The best-fit parameters and corresponding uncertainties are estimated using MCMC. We use the same fitting method to re-determine the MZR at $z\sim0$ (red line) and $z\sim0.7$ (pink line) for the \citet{zhuang_glimpse_2023} and \citet{beverage_carbon_2023} samples, respectively. We emphasize again that the individual metallicity measurements at $z\sim0$ and $z\sim0.7$ were made using the same full-spectrum modeling method as the $z=1-3$ sample. 

The best-fit stellar MZRs at each redshift interval correspond to the following relations:
\\

\noindent $z=1-3:$
$$\log\left(Z/Z_\odot\right)=0.19_{-0.12}^{+0.17} \;\log\left(\frac{M_*}{10^{11}\,M_\odot}\right) +0.13_{-0.11}^{+0.05}$$

\noindent $z\sim0.7:$
$$\log\left(Z/Z_\odot\right)=0.08_{-0.04}^{+0.04} \;\log\left(\frac{M_*}{10^{11}\,M_\odot}\right) +0.21_{-0.02}^{+0.02}$$

\noindent $z\sim0:$
$$\log\left(Z/Z_\odot\right)=0.27_{-0.02}^{+0.02} \;\log\left(\frac{M_*}{10^{11}\,M_\odot}\right) +0.30_{-0.02}^{+0.02}$$

The slope of the MZR at $z=1-3$ is consistent with that at $z\sim0$, within the uncertainties. The normalization, though, appears to have increased by $0.17\pm0.09$ since $z\sim2$. We note that the sample at $z=1-3$ is still small, and the metallicity measurements have significant statistical uncertainties. Consequently, the evidence for evolution remains below the $2\sigma$ level. Thus, larger samples and deeper observations are needed to confirm this potential evolution with more certainty.


An evolving MZR, if confirmed, is not surprising given that the total metal content of a galaxy is primarily in $\alpha$ elements (70 percent\footnote{Calculated using the \citet{asplund_chemical_2009} solar abundances and taking $\alpha$ elements to include N, O, Mg, Na, Ne, S, Si, Ti, and Ca.} by mass) and that we find slightly lower Mg in Figure~\ref{fig:abundance_pattern}. Even though distant quiescent galaxies are Fe deficient, Fe-peak elements only contribute 10 percent by mass to the total metallicity. Therefore, the observed offset in the MZR to lower $Z/Z_\odot$ could be explained by these galaxies having lower Mg abundances. 

Other studies based on Lick indices instead point to a redshift-invariant MZR, with galaxies at $z\sim2$ having super-solar $Z_*$ \citep{lonoce_old_2015, onodera_ages_2015}. However, these conclusions are based on individual measurements or a single stacked spectrum, and therefore they carry large uncertainties. Similarly, \citet{estrada-carpenter_clear_2019} report a redshift-invariant MZR, with stellar metallicities derived from low-resolution spectroscopy. In section~\ref{sec:prospector}, we compare these methods and demonstrate that metallicities derived from spectrophotometric fitting primarily reflect [Fe/H]. Given the Fe deficiencies found in this work, it is notable that \citet{estrada-carpenter_clear_2019} find no evolution in the MZR. 


\section{Discussion \& Implications}
\label{sec:discuss}

\subsection{Star-formation timescales}
\label{sec:timescale}

In this work, we present the first abundance pattern study (including Fe, C, Mg, Si, Ca, Ti) based on a sample of $z>1$ quiescent galaxies. \citet{jafariyazani_resolved_2020} had previously measured an abundance pattern at $z\sim2$, but only for a single lensed system, whereas \citet{beverage_heavy_2024} measured only the Mg and Fe abundances of a larger sample of quiescent galaxies at $z\sim1.4$ and $z\sim2.1$. In this section, we discuss the implication of multi-element abundance results on the star-formation histories of distant quiescent galaxies. 

A key observation from Figure~\ref{fig:abundance_pattern} was that massive queiscent galaxies at $z=1-3$ are deficient in Fe and C, whereas Mg, Si, Ca, and Ti have higher abundances. When considering the full ISM enrichment over a stellar population's lifetime, Fe is forged approximately equally in CC and SNe Ia, whereas C is enriched approximately equally by CC SNe and low-mass AGB stars. The other elements, Mg, Si, Ca, and Ti, are all primarily $\alpha$ elements, produced in CC SNe. These processes enrich over a diverse set of timescales; exploding massive stars (CC SNe) eject $\alpha$ elements almost instantaneously after the commencement of star formation, swiftly incorporating them into subsequent stellar generations. Alternatively, low- to intermediate-mass stars enrich the ISM on a delayed timescale due to their longer lifespans. One mechanism is via SNe Ia, which typically eject Fe-peak elements (e.g., Cr and Fe) only after a time delay of $\sim0.5-1$ Gyr. The other mechanism is via AGB stars, which have even shorter delay times than SNe Ia, enriching primarily C (and N) as early as 50~Myr but with a typical delay of $\sim250$~Myr \citep[e.g.,][]{cristallo_evolution_2011, cristallo_evolution_2015, maiolino_re_2019, johnson_empirical_2023}. Thus, galaxies that stop forming stars before significant AGB and SNe Ia contributions have very low Fe-peak and C$+$N stellar abundances, but still high $\alpha$ abundances.

Considering the different enrichment timescales, our low C and Fe abundance results imply that the SUSPENSE galaxies formed most of their stellar mass before significant AGB and Ia SNe enrichment, corresponding to a timescale of $\lesssim0.2$~Gyr. Such an extreme star-formation timescale would translate to an SFR of $\geq500~M_\odot\,{\rm yr}^{-1}$, putting these galaxies among the most vigorous star-forming galaxies in the Universe \citep[e.g.,][]{riechers_dust-obscured_2013,decarli_rapidly_2017,gullberg_alma_2019,liao__alma_2024}. 

This interpretation assumes that the observed deficiencies in C and Fe are solely driven by shorter star-formation timescales in higher-redshift galaxies. However, variations in the IMF may also play a significant role in shaping the relative elemental abundances. For instance, a top-heavy IMF would increase the relative number of CC to Ia supernovae, increasing the Mg producers over the Fe producers. This would elevate [Mg/Fe] without requiring extremely short star-formation timescales \citep[e.g.,][]{matteucci_abundance_1994,thomas_constraints_1999}. Furthermore, a top-heavy IMF could alter the IMF-averaged CC SNe yields, introducing even more variation. A top-heavy IMF is indeed plausible in these extreme galaxies, potentially due to unique environmental conditions or the effect of an integrated galaxy-wide IMF \citep[e.g.,][]{fontanot_variations_2017}. However, the extent of this impact---particularly on elements that are produced by multiple enrichment pathways---remains unclear. Detailed chemical evolution modeling is necessary to disentangle the respective roles of the IMF and star-formation timescales in producing these unusual abundance patterns.

Such detailed chemical evolution models would also provide more precise estimates of star-formation timescales, a task we leave for a future work. However, we note that the inferred extreme star-formation timescales conflict with the results from spectrophotometric fitting using the non-parametric star-formation histories of the SUSPENSE galaxies \citet{slob_jwst-suspense_2024}. In Sections~\ref{sec:prospector} and \ref{sec:impossible} we discuss the implications of these differences.

\subsection{The assembly of massive galaxies over cosmic time}
\label{sec:assembly}

Our elemental abundance patterns imply that galaxies at $z\sim2$ on average formed earlier (Section~\ref{sec:results}) and over shorter timescales (Section~\ref{sec:timescale}) compared to galaxies with similar velocity dispersions at $z\sim0$ and $z\sim0.7$. The most straightforward explanation for this observed increase is that galaxies in the quiescent sample at $z=1-3$ are among the earliest quenchers in the $z\sim0$ population; galaxies that form over longer timescales quench and join the quiescent galaxy population at later times. Thus, by $z=0$, the SUSPENSE sample represents only the extreme tail of the $t_\mathrm{form}$ distribution. This explanation is akin to the ``progenitor bias'' scenario \citep[i.e.,][]{van_dokkum_morphological_2001} used, for example, to explain the size growth of quiescent galaxies over time \citep[e.g.,][]{khochfar_simple_2006, van_dokkum_growth_2010, carollo_newly_2013, poggianti_evolution_2013}.

This progenitor bias scenario is reinforced by Figure~\ref{fig:formation_time-label}, where we find a negative correlation between formation time and [Mg/Fe], irrespective of the observed redshift of the galaxies. Thus, the evolution in C and Fe is mostly due to the lower redshift samples naturally including galaxies with earlier formation times and shorter star-formation timescales. 

A closer look at Figure~\ref{fig:formation_time-label}, however, reveals that progenitor bias alone may not be able to explain the observations. If the $z\sim2$ galaxies passively evolve over the past 10 billion years, without merging or experiencing late-time star formation, then the trends with formation time at $z\sim0$ and $z\sim2$ should exactly overlap at the earliest epochs. Instead, we find very few SDSS galaxies at these early formation times and we find that the $z\sim2$ samples exhibit slightly lower [Fe/H] and [Mg/H], and higher [Mg/Fe] than the lower redshift samples. This offset was first identified by \citet{zhuang_glimpse_2023} using a small sample of objects at $z=1-3$, and later confirmed by \citet{beverage_heavy_2024} with a larger sample.


\begin{figure*}
    \centering
    \includegraphics[width=\textwidth]{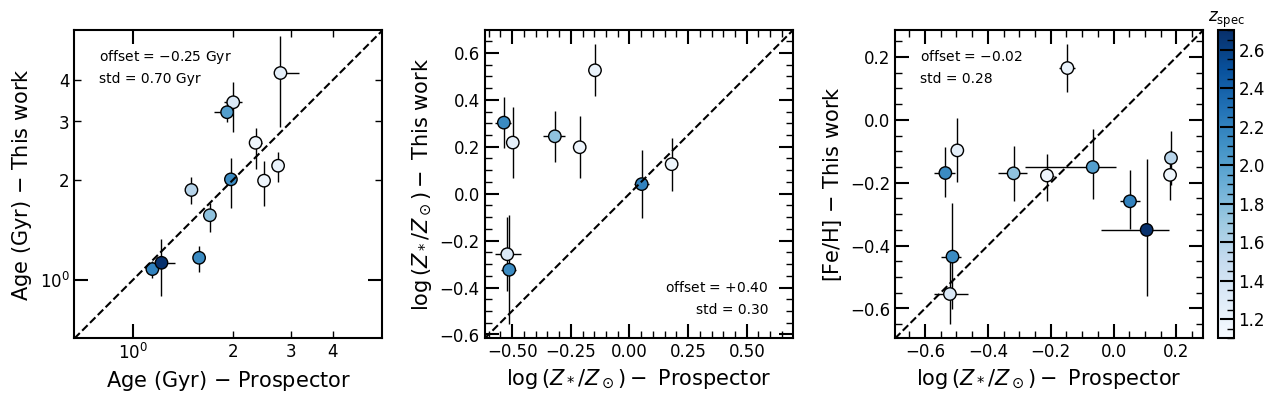}
    \caption{Comparison between \textsc{Prospector} and \texttt{alf} measurements of the ages (left), metallicities (middle), and [Fe/H] (right) of the massive quiescent galaxies at $z=1-3$ from \textit{JWST}-SUSPENSE. The dashed black line shows a one-to-one relation. The data points are colored by their spectroscopic redshifts. The metallicities from this work are calculated using the \citep{thomas_stellar_2003} conversion to [Z/H] from [Fe/H] and [Mg/Fe]. The standard deviation of the points and mean offset are listed in each panel.}
    \label{fig:prospector}
\end{figure*}

One way to explain the lack of chemically extreme galaxies at $z\sim0$ is late-time star formation episodes. If the star-formation material is pre-enriched with SN Ia products by previous epochs of star formation, then the newly formed stars would be younger, more metal-rich, and more $\alpha$/Fe enhanced, pushing them along the abundance trends of Figure~\ref{fig:formation_time-label} towards younger ages (later formation times). Major mergers between galaxies with different star-formation histories would also shift galaxies upward along the formation time sequence. In contrast, minor mergers tend to decrease not only [Fe/H] but also [Mg/H] levels. However, due to aperture effects, interpreting the impact of minor mergers is more complicated because they are preferentially accreted to the outskirts of galaxies. Therefore, minor mergers likely have minimal influence on the observed abundance patterns. Larger sample sizes at $z\gtrsim1$ and $z\sim0$ with careful consideration of stellar mass completeness is required to clarify the relative importance of progenitor bias, mergers, and late-time star formation.

\subsection{Galaxy quenching}
\label{sec:quench}

A key finding from Figure~\ref{fig:formation_time-label} is galaxies that form at earlier epochs have lower [Mg/H]. Unlike the correlation with [Fe/H], this result cannot be explained using star-formation timescales, because [Mg/H] instead reflects instantaneous enrichment by CC SNe. 

One possible explanation for galaxies with earlier formation times having slightly lower [Mg/H] is that they expel larger gas reservoirs during the quenching phase. \citet{beverage_elemental_2021} proposed this model to explain a similar trend in the LEGA-C sample. They used a leaky box model with exponentially declining inflow rates and SNe-driven outflows, with analytical solutions from \citet{spitoni_new_2017}. In their models, galaxies with smooth star-formation histories that quench via slow gas depletion end up with high stellar [Mg/H] by $z=0$, regardless of their star-formation history.

The implementation of rapid gas expulsion after two $e$-folding times (akin to AGN-driven feedback), successfully keeps the stellar [Mg/H] low. In these models, galaxies that quenched at higher SFR and thus expel larger gas reservoirs, have lower [Mg/H] and earlier formation times. Thus, the observed correlation between [Mg/H] and $t_\mathrm{form}$ may indicate more efficient gas expulsion at higher redshift. These more efficient outflows in combination with progenitor bias, in turn, can explain the increase in [Mg/H] \textit{and} the MZR over cosmic time.

This result, in combination with the extreme SFRs inferred from the C and Fe deficiencies, is consistent with quasar-driven quenching resulting from rapid gas inflows. Indeed, evidence of ejective AGN outflows has been found in galaxies that rapidly quenched after $z\sim2$, indicated by pronounced blueshifted Na~\textsc{I} D and other ISM absorption lines \citep[e.g.,][]{maltby_high-velocity_2019, Kubo2022, belli_massive_2023, deugenio_fast-rotator_2023, davies_jwst_2024, park_widespread_2024}. Future chemical evolution modeling will help determine the mass-loading factors and star-formation timescales required to reproduce the trends with $t_\mathrm{form}$.

\subsection{SPS fitting}
\label{sec:prospector}

In this section, we compare measurements derived in this work with those obtained using a more standard stellar population modeling approach, which employs solar abundance patterns and non-parametric star-formation histories. \citet{slob_jwst-suspense_2024} present stellar population parameters for the SUSPENSE galaxies using the \textsc{Prospector} code \citep{leja_how_2019, johnson_stellar_2021}. \textsc{Prospector} utilizes the the Flexible Stellar Population Synthesis \citep[\texttt{FSPS};][]{conroy_propagation_2009,conroy_propagation_2010} library, the MILES spectral library, and the MIST isochrones, assuming a \citet{chabrier_galactic_2003} IMF. The fitting process incorporates both \textit{JWST} spectroscopy and UltraVISTA DR3 photometry. For more details, see \citet{slob_jwst-suspense_2024}.

There are a few key differences between this standard modeling approach and the method used in this work. Firstly, in \textsc{Prospector} and similar codes, such as \textsc{Bagpipes} \citep{carnall_inferring_2018}, \textsc{Cigale} \citep{burgarella_star_2005}, and \textsc{Magphys} \citep{da_cunha_simple_2008}, the star-formation history (SFH) is a free parameter. Secondly, they assume a solar-scaled abundance pattern (and sometimes $Z=Z_\odot$). Another key difference is that \textsc{Prospector}, when fitting both the photometry and spectra, heavily relies on the galaxy continuum shape and consequently does not put emphasis on reproducing the individual absorption features. In contrast, our full-spectrum modeling assumes a variable abundance pattern, a single burst of star formation, and relies solely on the stellar absorption features. 

Before comparing the results from these codes, it is important to highlight that differences in the SFH assumptions lead to the derived quantities being either light-weighted values in \textsc{Prospector} or SSP-equivalent values in \textsc{alf$\alpha$}. Thus, it is crucial to assess how these assumptions may impact the inferred properties. Recent work by \citet{gountanis_modeling_2024} tested the SSP-assumption by generating composite spectra using a superposition of SSPs with evolving ages, metallicities, and abundance ratios [Mg/Fe], informed by a delayed-tau star-formation history and chemical evolution models. They then fit these composite spectra using \texttt{alf$\alpha$}, assuming a single SSP. The results showed excellent agreement between the SSP-equivalent parameters and the true light-weighted averages, with discrepancies within 0.05 dex for ages, metallicities, and [Mg/Fe] (see their Figure 13). This test indicates that using an SSP-equivalent approach in \texttt{alf$\alpha$} does not introduce significant biases, allowing for a valid comparison between the codes.



In Figure~\ref{fig:prospector}, we compare our best-fit ages and metallicities to those from \textsc{Prospector}. The points are colored by their spectroscopic redshift, and the dashed black lines represent the one-to-one line. The left panel shows stellar ages, where \textsc{Prospector} provides mass-weighted estimates and our full-spectrum fitting gives SSP-equivalent ages. Despite different assumptions in star-formation history (extended non-parametric vs. single age), the two sets of stellar ages agree remarkably well. This result may not be surprising, as all of the galaxies in our sample have been quiescent for at least $\gtrsim1$ Gyr. At these older ages, these stellar populations are less susceptible to the outshining problem, wherein the youngest stars with low $M/L$ dominate the continuum.

Next, we examine stellar metallicities in the middle panel of Figure~\ref{fig:prospector}, comparing $\log\left(Z/Z_\odot\right)$. As a reminder, the total metallicities from this work were computed using a combination of the best-fit [Mg/Fe] and [Fe/H]. We find poor agreement between our total metallicities and those from \textsc{Prospector}, with significant scatter (std = 0.32 dex) and a systematic offset of 0.41 dex towards lower values in the \textsc{Prospector} measurements. When we compare the \textsc{Prospector} metallicities instead to [Fe/H], the systematic offset disappears. However, the large scatter around the one-to-one line remains. Additionally, the uncertainties on the \textsc{Prospector} measurements are significantly underestimated. 



The reason the \textsc{Prospector} $\log\left(Z/Z_\odot\right)$ aligns better with [Fe/H] rather than the total stellar metallicity is that Mg has less impact on the spectral shape compared to Fe \citep[e.g.,][]{choi_imprint_2019, leja_older_2019}. As discussed in Section~\ref{sec:mzr}, despite Fe's significant contribution to the absorption features in galaxies, it contributes only a small fraction to the total metal content. This effect is even more pronounced in light of the Fe-deficiencies found for the $z=1-3$ sample.

The significant disagreement between the metallicities derived from \textsc{Prospector} and those presented here is surprising, especially given that both methods use nearly identical stellar population models (MIST isochrones and MILES stellar libraries). However, three key differences remain: \textsc{Prospector} incorporates photometry (capturing continuum shape), allows for an extended SFH, and uses a solar-scaled abundance pattern. To investigate the source of the discrepancy, we conducted the following tests. First, running \textsc{Prospector} on the continuum-normalized spectrum (excluding photometry) still showed a significant discrepancy with the results presented here, indicating the continuum shape is not the main cause. Second, constraining \textsc{Prospector} to an SSP to match \textsc{alf$\alpha$}'s SFH did not resolve the disagreement either. Finally, additionally forcing \textsc{alf$\alpha$} to adopt a solar-scaled abundance pattern improved the metallicity agreement, though significant scatter remained. This final test was the only case where the metallicities from \textsc{Prospector} and \textsc{alf$\alpha$} showed significant correlation according to the Pearson correlation test (p-value$=0.08$). These results suggest that while multiple factors contribute, fitting non-solar populations with solar-scaled abundance patterns significantly impacts the metallicity estimates.

This impact of assuming a solar-scaled abundance pattern when fitting $\alpha$-enhanced galaxies is further demonstrated by the performance of \textsc{Prospector}'s best-fit models, as shown in \citet{slob_jwst-suspense_2024}. \textsc{Prospector} often struggles to reproduce key spectral features, particularly those associated with Mg, Si, and C, underscoring that solar-scaled models fail to capture the non-solar abundances present in these galaxies. This finding is corroborated by the reduced $\chi^2$ values from fitting the spectra with solar-scaled abundance patterns with \textsc{alf$\alpha$}: solar-scaled fits result in a typical reduced $\chi^2$ of 1.3, while allowing for non-solar abundances improves the fit to 1.1.




Therefore, caution is needed when interpreting or assuming metallicities derived from codes using a solar abundance pattern. Such measurements often carry underestimated uncertainty, fail to trace the true metallicity, and, on average, are more likely to reflect the Fe abundance, which is a biased indicator of total metallicity.

Finally, we compare the best-fit SFHs from \textsc{Prospector} to those inferred from the elemental abundances. The typical star-formation timescales from \textsc{Prospector}, which we define as $\left(t_{84}-t_{16}\right)$\footnote{where $t_{x}$ corresponds to the age of the universe at which $x$ percent of the stellar mass has been formed}, is 1.2 Gyr, a factor of 6$\times$ longer than that predicted by the C and Fe deficiencies and chemical evolution arguments ($\sim$200~Myr). It is perhaps not surprising that the SFHs derived from Prospector tend to be longer. At these stellar ages (1-4 Gyr), the time resolution of the SPS models is only $\sim1$\;Gyr \citep[][]{zibetti_maximum_2024}, since the spectra of older stellar populations are less sensitive to changes in age. In contrast, the spectra of young quiescent galaxies are significantly more sensitive to age, and age resolution can be as low as $\sim$20\;Myr \citep[][]{suess_recovering_2022}. This lower resolution in older stellar populations motivates future work that integrates stellar population modeling with priors informed by chemical evolution, which could help reveal the shorter star-formation timescales we infer from chemical evolution arguments. These shorter timescales provide crucial insights into the build-up of massive galaxies, as discussed in the next section.

\subsection{The early formation of massive galaxies}
\label{sec:impossible} 

The short star-formation timescales inferred from the abundance patterns may have broader implications for the formation of massive galaxies in the early universe. Recently, \textit{JWST} identified a population of quiescent and extremely massive galaxies at $z\gtrsim3$ \citep{antwi-danso_feniks_2023, de_graaff_efficient_2024, glazebrook_massive_2024,stawinski_spectroscopic_2024,carnall_jwst_2024}. The SFHs of these galaxies, derived using \textsc{Prospector} and similar fitting codes, indicate that they begin to accumulate their stellar mass at very early epochs ($z\gtrsim10$). It has been suggested that this early stellar mass growth may be in conflict with the predictions of $\Lambda$CDM, even when assuming a maximum (100\%) baryon-to-star conversion efficiency, highlighting the ``impossibly early galaxy problem'' \citep[e.g.,][]{steinhardt_impossibly_2016}.

One way to explain this tension is that the SFHs derived from standard modeling techniques are biased towards more extended SFHs. Most of the non-parametric SFHs were intentionally designed to include an early build-up of stellar mass, to solve inconsistencies with lower redshift galaxy evolution studies \citep{leja_how_2019, leja_beyond_2019, carnall_how_2019, carnall_jwst_2024}. Therefore, these models may not apply to the $z\gtrsim2$ galaxy population. \citet{carnall_jwst_2024} fitted an instantaneous burst of star formation to three massive quiescent galaxies at $z\gtrsim3$. In this test, the best-fit stellar ages were consistent with the original fits; however, the initial buildup of stellar mass was delayed. These results show that these early massive galaxies could be explained within the $\Lambda$CDM framework by simply shortening the assumed SFHs.


Our elemental abudance results indeed show that standard stellar population modeling codes overestimate the duration of the star-forming phase. Thus, assuming that quiescent galaxies at $z>3$ will have similarly short star formation timescales, the early mass buildup in these massive galaxies may be less problematic than found by earlier studies. As elemental abundance studies will be prohibitively difficult at $z>3$, our elemental abundance pattern may help set more informative priors on the SFHs in stellar population modeling codes.



\section{Summary}
\label{sec:summary}

In this Paper, we present the stellar metallicities and multi-element abundances (C, Mg, Si, Ca, Ti, Cr, and Fe) of 15 massive quiescent galaxies at $z=1-3$ from the \textit{JWST}-SUSPENSE program. The ultradeep (16 hour) NIRSpec/MSA spectra were modeled using the custom full-spectrum fitting code \texttt{alf$\alpha$}, a Python implementation of \texttt{alf} \citep[]{conroy_metal-rich_2018}, which has been made publicly available. 

Compared to the $z\sim0$ and $z\sim0.7$ population of massive quiescent galaxies, those at $z=1-3$ have $-$0.26$\pm0.04$~dex lower [C/H], $-0.16\pm0.03$~dex lower [Fe/H], and $-0.07\pm0.04$~dex lower [Mg/H]. There is also evidence that the quiescent MZR at $z=1-3$ is lower by $0.17\pm0.09$~dex, but the uncertainties are large and the sample size is small. The C and Fe deficiencies indicate that distant quiescent galaxies form over shorter star-formation timescales than today's quiescent galaxy population, quenching before significant enrichment by AGB stars and SNe Ia. Such rapid star-formation timescales ($\sim0.2$ Gyr) correspond to extreme star-formation rates of $\sim500~M_\odot\,{\rm yr}^{-1}$ at $z\sim2-10$, putting these galaxies among the most vigorous star-forming galaxies in the Universe. In the future, chemical evolution modeling will offer more precise constraints on the star-formation timescale, while also providing insights into the role of a top-heavy IMF in shaping this timescale.

We also find correlations between galaxy formation time and [Fe/H], [Mg/H], and [Mg/Fe], such that galaxies that form at earlier times have abundances consistent with shorter star-formation timescales (i.e., higher [Mg/Fe] and lower [Fe/H]). These trends had previously been found within the $z\sim0$ and $z\sim0.7$ populations \citep[][]{zhuang_glimpse_2023, beverage_carbon_2023}{}{}, however, using the $z=1-3$ results, we show that the same trends extend to higher redshift and earlier formation times. This result suggests that the observed evolution in [C/H] and [Fe/H] over cosmic time is driven by lower redshift samples naturally including galaxies that formed over longer timescales. In other words, the $z=1-3$ quiescent galaxies represent the extreme tail of today's massive quiescent galaxy population. Interestingly, the $z\sim0$ sample lacks the chemically extreme galaxies at $z=1-3$, indicating mergers and/or late-time star formation likely contribute to the evolution in the elemental abundances. Larger sample sizes and careful consideration of completeness are required to clarify this picture.

Additionally, we confirm the marginal correlation between [Mg/H] and formation time, suggested by previous results with smaller sample sizes and larger measurement uncertainties \citep[][]{zhuang_glimpse_2023, beverage_heavy_2024}{}{}. Given that Mg is a tracer of instantaneous metal enrichment and not the star-formation timescale, this marginal trend may imply that galaxies that form at earlier times expel larger gas reservoirs during the quenching phase, as suggested previously by \citet{beverage_elemental_2021}. Combined with the extreme SFRs inferred from the C and Fe deficiencies, this interpretation is consistent with quenching by AGN-driven outflows. In the future, we will use chemical evolution modeling to measure the mass-loading factors and star-formation timescales required to reproduce the trends with $t_\mathrm{form}$. 

Next, we compare our stellar ages and metallicities to results from the spectrophotometric modeling code \textsc{Prospector}, which assumes a solar-scaled elemental abundance pattern and a non-parametric SFH. The stellar ages agree remarkably well, but the stellar metallicities disagree significantly. Furthermore, the \textsc{Prospector} metallicities carry vastly underestimated uncertainties. However, despite the large scatter, the \textsc{Prospector} metallicities are in better agreement with [Fe/H], \textit{not} the total metal content [Z/H]. We attribute this result to solar-scaled models being more sensitive to [Fe/H] because of the strong impact of Fe on the stellar spectrum. Thus, although Fe only contributes approximately 10 percent by mass to the total metal content of quiescent galaxies, its abundance significantly impacts the optical stellar spectrum. In light of the observed Fe deficiencies and underestimated measurement uncertainties, we therefore emphasize caution when interpreting or assuming metallicities from modeling codes that adopt solar abundance patterns.

Finally, we find that the star-formation timescales indicated by the extreme elemental abundance patterns of distant quiescent galaxies are significantly shorter than those predicted by standard spectrophotometric modeling codes. This discrepancy confirms that the SFHs from these codes may be overly biased towards extended and early stellar mass buildup when applied to distant quiescent galaxies \citep[e.g.,][]{carnall_how_2019,leja_how_2019,leja_beyond_2019}{}{}. Addressing this bias could help mitigate the possible tension with $\Lambda$CDM for quiescent galaxies at $z>3$ \citep[see also][]{carnall_jwst_2024}. As elemental abundance measurements become prohibitively challenging at $z>3$, these findings demonstrate how our elemental abundance patterns can provide more informative priors on the SFHs in stellar population modeling codes.

In this Paper, we demonstrate the power of \textit{JWST} for studying the multi-element abundances of distant quiescent galaxies. In the future, we will combine this expanding multi-element dataset at $z\gtrsim1$ with chemical evolution modeling, to uncover a more detailed picture of the star-formation histories, quenching, and assembly of massive quiescent galaxies over cosmic time. \\

This work is based on observations made with the NASA/ESA/CSA \textit{James Webb Space Telescope}. The data were obtained from the Mikulski Archive for Space Telescopes at the Space Telescope Science Institute, which is operated by the Association of Universities for Research in Astronomy, Inc., under NASA contract NAS 5-03127 for \textit{JWST}. These observations are associated with program JWST-GO-2110. Support for program JWST-GO-2110 was provided by NASA through a grant from the Space Telescope Science Institute, which is operated by the Association of Universities for Research in Astronomy, Inc., under NASA contract NAS 5-03127. AGB is supported by the National Science Foundation Graduate Research Fellowship Program under grant Nos. DGE 1752814 and DGE 2146752 and by the H2H8 Association and Leids Kerkhoven-Bosscha Fonds. MK acknowledges funding from the Dutch Research Council (NWO) through the award of the Vici grant VI.C.222.047 (project 2010007169). CC acknowledges support from NSF Award 1908748. 

%

\vspace{5mm}
\facilities{JWST(NIRSpec)}

The JWST data presented in this article were obtained from the Mikulski Archive for Space Telescopes (MAST) at the Space Telescope Science Institute. The specific observations analyzed can be accessed via \dataset[doi: 10.17909/6wjp-qb35]{https://doi.org/10.17909/6wjp-qb35}


\software{\textsc{Prospector}\citep{leja_how_2019,johnson_stellar_2021}, \textsc{alf}\citep{conroy_counting_2012,conroy_metal-rich_2018}, \textsc{alf$\alpha$}
          }



\bibliography{suspense_abundances}{}
\bibliographystyle{aasjournal}



\appendix

\section{Tables of elemental abundance ratios}
\label{app:table}

In this Appendix, we present tables of the abundance ratios [X/Mg] (Table~\ref{table:alf-results_xmg}) and [X/Fe] (Table~\ref{table:alf-results_xfe}) of the SUSPENSE galaxies. The uncertainties are derived from the MCMC chains, which account for potential non-Gaussianity in the posterior distributions of the absolute abundances, [X/H].

\begin{deluxetable*}{ccccccc}
\tabletypesize{\footnotesize}\tablecolumns{7}
\tablewidth{0pt}
\tablecaption{\label{table:alf-results_xmg}Elemental abundance ratios with Mg as the reference element}
\tablehead{\colhead{ID}\ & \colhead{[Fe/Mg]} \ & \colhead{[C/Mg]}\ & \colhead{[Ca/Mg]}\ & \colhead{[Ti/Mg]}\ & \colhead{[Cr/Mg]}\ & \colhead{[Si/Mg]}} 
\startdata
127345 & $-0.32^{+0.09}_{-0.09}$ & $-0.17^{+0.11}_{-0.12}$ & $-0.00^{+0.10}_{-0.10}$ & -- & $-0.33^{+0.13}_{-0.12}$ & $-0.24^{+0.15}_{-0.16}$\\ 
130040 & $-0.40^{+0.12}_{-0.11}$ & $-0.16^{+0.16}_{-0.17}$ & $-0.26^{+0.12}_{-0.12}$ & $0.15^{+0.18}_{-0.20}$ & $-0.19^{+0.16}_{-0.16}$ & $-0.29^{+0.18}_{-0.17}$\\ 
127154 & $-0.38^{+0.08}_{-0.08}$ & $-0.26^{+0.11}_{-0.11}$ & -- & $-0.08^{+0.14}_{-0.16}$ & $-0.49^{+0.14}_{-0.14}$ & $-0.17^{+0.14}_{-0.14}$\\ 
129982 & $-0.34^{+0.12}_{-0.12}$ & $-0.29^{+0.18}_{-0.17}$ & $-0.21^{+0.16}_{-0.17}$ & $-0.18^{+0.20}_{-0.19}$ & $-0.04^{+0.16}_{-0.16}$ & --\\ 
127108 & $-0.32^{+0.13}_{-0.13}$ & $-0.54^{+0.18}_{-0.17}$ & $-0.36^{+0.21}_{-0.21}$ & -- & $-0.70^{+0.18}_{-0.18}$ & $0.01^{+0.23}_{-0.28}$\\ 
129197 & -- & -- & -- & -- & -- & --\\ 
129149 & -- & -- & -- & -- & -- & --\\ 
128041 & $-0.44^{+0.07}_{-0.07}$ & $-0.45^{+0.09}_{-0.09}$ & $-0.39^{+0.07}_{-0.07}$ & $-0.11^{+0.12}_{-0.13}$ & $-0.54^{+0.11}_{-0.11}$ & $-0.49^{+0.14}_{-0.14}$\\ 
127700 & -- & -- & -- & -- & -- & --\\ 
129133 & $-0.50^{+0.08}_{-0.08}$ & $-0.32^{+0.09}_{-0.09}$ & $-0.11^{+0.09}_{-0.08}$ & -- & $-0.47^{+0.14}_{-0.13}$ & $-0.26^{+0.17}_{-0.18}$\\ 
127941 & $-0.12^{+0.16}_{-0.18}$ & $-0.16^{+0.19}_{-0.20}$ & $0.51^{+0.16}_{-0.17}$ & -- & $0.39^{+0.20}_{-0.22}$ & $0.32^{+0.24}_{-0.27}$\\ 
128036 & $-0.32^{+0.12}_{-0.12}$ & $-0.33^{+0.13}_{-0.13}$ & $0.09^{+0.12}_{-0.12}$ & -- & $-0.51^{+0.19}_{-0.18}$ & $-0.31^{+0.22}_{-0.23}$\\ 
128913 & -- & -- & -- & -- & -- & --\\ 
130725 & -- & -- & -- & -- & -- & --\\ 
129966 & -- & -- & -- & -- & -- & --\\ 
\enddata
\vspace{-0.25cm}
\vspace{-0.5cm}
\end{deluxetable*}

\begin{deluxetable*}{ccccccc}
\tabletypesize{\footnotesize}\tablecolumns{7}
\tablewidth{0pt}
\tablecaption{\label{table:alf-results_xfe}Elemental abundance ratios with Fe as the reference element}
\tablehead{\colhead{ID}\ & \colhead{[Mg/Fe]} \ & \colhead{[C/Fe]}\ & \colhead{[Ca/Fe]}\ & \colhead{[Ti/Fe]}\ & \colhead{[Cr/Fe]}\ & \colhead{[Si/Fe]}} 
\startdata
127345 & $0.32^{+0.09}_{-0.09}$ & $0.15^{+0.10}_{-0.10}$ & $0.32^{+0.09}_{-0.09}$ & -- & $-0.01^{+0.12}_{-0.12}$ & $0.08^{+0.13}_{-0.14}$\\ 
130040 & $0.40^{+0.11}_{-0.12}$ & $0.24^{+0.13}_{-0.14}$ & $0.14^{+0.10}_{-0.10}$ & $0.55^{+0.20}_{-0.23}$ & $0.21^{+0.16}_{-0.16}$ & $0.11^{+0.15}_{-0.14}$\\ 
127154 & $0.38^{+0.08}_{-0.08}$ & $0.13^{+0.09}_{-0.10}$ & -- & $0.31^{+0.14}_{-0.17}$ & $-0.10^{+0.13}_{-0.14}$ & $0.22^{+0.12}_{-0.13}$\\ 
129982 & $0.34^{+0.12}_{-0.12}$ & $0.05^{+0.16}_{-0.16}$ & $0.13^{+0.15}_{-0.15}$ & $0.16^{+0.21}_{-0.22}$ & $0.29^{+0.16}_{-0.16}$ & --\\ 
127108 & $0.32^{+0.13}_{-0.13}$ & $-0.22^{+0.14}_{-0.14}$ & $-0.05^{+0.20}_{-0.19}$ & -- & $-0.40^{+0.15}_{-0.14}$ & $0.32^{+0.19}_{-0.24}$\\ 
129197 & -- & -- & -- & -- & -- & --\\ 
129149 & -- & $0.13^{+0.10}_{-0.09}$ & $-0.07^{+0.08}_{-0.08}$ & $0.35^{+0.16}_{-0.18}$ & $0.18^{+0.16}_{-0.15}$ & $0.12^{+0.16}_{-0.18}$\\ 
128041 & $0.44^{+0.07}_{-0.07}$ & $-0.01^{+0.08}_{-0.08}$ & $0.05^{+0.06}_{-0.06}$ & $0.34^{+0.12}_{-0.13}$ & $-0.10^{+0.11}_{-0.11}$ & $-0.05^{+0.12}_{-0.13}$\\ 
127700 & -- & $-0.08^{+0.15}_{-0.15}$ & $-0.02^{+0.13}_{-0.12}$ & -- & -- & --\\ 
129133 & $0.50^{+0.08}_{-0.08}$ & $0.18^{+0.09}_{-0.09}$ & $0.39^{+0.08}_{-0.07}$ & -- & $0.03^{+0.14}_{-0.13}$ & $0.24^{+0.16}_{-0.17}$\\ 
127941 & $0.12^{+0.18}_{-0.16}$ & $-0.04^{+0.17}_{-0.17}$ & $0.62^{+0.12}_{-0.12}$ & -- & $0.51^{+0.18}_{-0.21}$ & $0.44^{+0.21}_{-0.24}$\\ 
128036 & $0.32^{+0.12}_{-0.12}$ & $-0.01^{+0.11}_{-0.12}$ & $0.41^{+0.09}_{-0.09}$ & -- & $-0.19^{+0.19}_{-0.18}$ & $0.01^{+0.19}_{-0.21}$\\ 
128913 & -- & -- & -- & -- & -- & --\\ 
130725 & -- & -- & $0.26^{+0.20}_{-0.20}$ & -- & -- & --\\ 
129966 & -- & -- & -- & -- & -- & --\\ 
\enddata
\vspace{-0.25cm}
\vspace{-0.5cm}
\end{deluxetable*}





\end{document}